\documentclass[aps,prd,amsmath,amsfonts,a4paper,11pt,reprint,preprintnumbers,twocolumn]{revtex4-1}
\usepackage{hyperref}
\usepackage{natbib}
\usepackage{amsmath}
\usepackage{amsfonts}
\usepackage{amssymb}
\usepackage{bm}

\newcommand{\grad}{\nabla}
\newcommand{\e}[1]{\mathrm{e}^{{#1}}}
\renewcommand{\d}{\mathrm{d}}
\newcommand{\vect}[1]{\bm{\mathrm{{#1}}}}
\DeclareMathOperator{\Si}{Si}
\DeclareMathOperator{\Ci}{Ci}
\DeclareMathOperator{\Ei}{Ei}
\newcommand{\im}{\mathrm{i}}
\newcommand{\EulerGamma}{\gamma_{\mathrm{E}}}
\DeclareMathOperator{\Or}{O}
\renewcommand{\leq}{\leqslant}

\newcommand{\D}{\mathrm{D}}

\newcommand{\Cconst}{\omega}

\newcommand{\Mp}{M_{\mathrm{P}}}
\newcommand{\Lag}{\mathcal{L}}
\newcommand{\LagM}{\Lag_{\mathrm{M}}}
\newcommand{\Gammaloop}{\Gamma_{\mathrm{q}}}

\newcommand{\fnl}{f_{\mathrm{NL}}}
\newcommand{\gnl}{g_{\mathrm{NL}}}
\newcommand{\tnl}{\tau_{\mathrm{NL}}}
\newcommand{\deltaG}{\delta_{\mathrm{g}}}
\newcommand{\Ps}{\mathcal{P}}

\newcommand{\etal}{\emph{et al.}}
\newcommand{\para}[1]{\par\vspace{1mm}\noindent\textbf{{#1}.}}
\newcommand{\pt}[1]{(\emph{{#1}})}

\begin{document}

	\title{Galileon inflation}
	\preprint{DESY 10-132}

	\author{Clare Burrage}
	\affiliation{Theory Group, Deutsches Elektronen-Synchrotron DESY,
	D-22603, Hamburg, Germany}
	\email{clare.burrage@desy.de}

	\author{Claudia \surname{de Rham}}
	\affiliation{D\'{e}partment de Physique Th\'{e}orique,
	Universit\'{e} de Gen\`{e}ve, 24 Quai E. Ansermet,
	CH-1211, Gen\`{e}ve, Switzerland}
	\email{Claudia.deRham@unige.ch}

	\author{David Seery}
	\affiliation{Department of Physics and Astronomy, University of Sussex,
	Brighton, BN1 9QH, UK}
	\email{D.Seery@sussex.ac.uk}

	\author{Andrew J. Tolley}
	\affiliation{Department of Physics, Case Western Reserve University,
	10900 Euclid Avenue, Cleveland OH, 44106-7079}
	\email{andrew.j.tolley@case.edu}

	\begin{abstract}
	Galileon inflation is a radiatively stable higher derivative
	model of inflation. The model is determined by a finite number
	of relevant operators which are protected by a covariant
	generalization of the Galileon shift symmetry.	We show that
	the nongaussianity of the primordial density perturbation
	generated during an epoch of Galileon inflation is
	a particularly powerful observational probe of these models
	and that, when the speed of sound is small, $\fnl$ can be larger than
	the usual result $f_{NL}\propto c_s^{-2}$.
	\end{abstract}

	\maketitle

	\section{Introduction}
	\label{sec:intro}

	Inflation is an era during which the cosmological scale
	factor $a(t)$ satisfies $\d^2 a / \d t^2 > 0$ as a function of
	cosmic time $t$, growing by a factor $\e{N}$.
	It is a familiar idea that a successful implementation of
	inflation, by which we mean
	obtaining sufficiently large $N$,
	requires the inflationary Lagrangian density $\Lag$ to have an approximate
	shift symmetry---an invariance under the translation
	$\phi \rightarrow \delta_c \phi \equiv
	\phi + c$, where $\phi$ is the inflaton field and
	$c$ is a constant.
	
	Underlying this idea
	is the observation that slow-roll
	implies the shift symmetry is broken only mildly,
	both in the action and the equations of motion.
	In terms of the inflationary potential, $V(\phi)$,
	and the Planck mass, $\Mp$,
	these
	slow-roll conditions are typically
	expressed using the parameters $\epsilon$
	and $\eta$, which satisfy $2 \epsilon \equiv \Mp^2 (V'/V)^2$
	and $\eta \equiv \Mp^2 V''/V$.
	
	How large a breaking can be acceptable?
	We expect any effective description to be valid only for
	a field excursion at most of order $\Mp$,
	before renormalization group flow introduces new physics which changes
	the	description.
	Requiring the first- and second-order fractional
	variation in the potential, $V(\phi)$, to be small over
	an excursion of this magnitude yields
	\begin{equation}
		\delta_c \ln V \sim \sqrt{\epsilon} \ll 1 ,
		\quad \text{and} \quad
		\delta_c^2 \ln V \sim \eta - \epsilon \ll 1 ,
		\label{eq:sr-break}
	\end{equation}
	where we have indicated approximate relations
	in terms of the slow-roll quantities $\epsilon$ and
	$\eta$. Therefore slow-roll inflation, defined by the
	conditions $\epsilon \ll 1$ and $|\eta| \ll 1$,
	entails tuning $V(\phi)$
	so that only very mild breaking occurs even over large
	variations in field value.
	
	This tuning has an important consequence.
	Creminelli observed that,
	once $V(\phi)$ has been adjusted to satisfy Eq.~\eqref{eq:sr-break},
	we can add
	any operator invariant under the
	shift symmetry without spoiling the property of successful
	inflation \cite{Creminelli:2003iq}.
	There is a large class of such operators, constructed by applying
	any combination of derivatives to the inflaton field $\phi$.
	This yields $\grad \phi$, $\grad \grad \phi$
	and higher gradients, with indices contracted in arbitrary
	combinations.
	It follows that the most general local,
	diffeomorphism-invariant
	action for $\phi$
	coupled to
	Einstein
	gravity which is invariant under
	the shift symmetry can be written
		\footnote{Although not
		written explicitly, the
		field $\phi$ may be non-minimally coupled to
		gravity. (If desired, this could
		be arranged using commutators of the
		derivative $\grad$.) Likewise, couplings to other
		geometric quantities may be present, such as the
		Gauss--Bonnet invariant $G$, for which a term of the form
		$\phi G$ would not spoil the shift symmetry
		and has second-order equations of motion.
		For simplicity we suppress
		such terms in Eq.~\eqref{eq:action}.
		More generally, as we will
		explain in \S\ref{sec:gal-inflation},
		although present in principle,
		they would contribute operators of higher dimension
		than those we keep, and can therefore be neglected.}
	\begin{equation}
		\Lag = \sqrt{-g} \left[
			\frac{\Mp^2}{2} R
			+ \LagM( \grad \phi, \grad \grad \phi, \ldots )
			\right] .
		\label{eq:action}
	\end{equation}
	
	In this paper, we study how inflation can be realized
	in theories of the form~\eqref{eq:action}.
	In comparison with slow-roll inflation using canonical kinetic terms
	there are new difficulties, associated with the appearance of unstable
	`ghost' states and stability under radiative corrections.
	One class of ghost-free, radiatively stable models with an interesting
	inflationary phenomenology has been widely studied.
	These are the Dirac--Born--Infeld (DBI) models
	\cite{Silverstein:2003hf, Chen:2005ad, Alishahiha:2004eh}.
	However, another class of ghost-free models
	has recently been constructed
	\cite{Nicolis:2008in, deRham:2010eu},
	based on a so-called `Galilean' symmetry to be defined
	in \S\ref{sec:inflation} below.
	This Galileon field has been shown
	to arise naturally in theories of massive gravity without ghosts in
	their decoupling limit \cite{deRham:2010gu, *deRham:2010ik}.

	Galileon models describe an effective \emph{short}-distance
	theory associated with a modification of gravity
	on large scales, and have mostly
	received attention
	as models of dark energy.
	However,
	it is equally possible that they
	can be used to describe cosmological evolution,
	and have been investigated for this purpose by numerous
	authors
	\cite{Chow:2009fm, *Silva:2009km, *Kobayashi:2009wr,
	*Kobayashi:2010wa,
	*DeFelice:2010pv, *DeFelice:2010nf}.
	Ali {\etal} obtained constraints on the parameters of the model
	(to be discussed in \S\ref{sec:gal-inflation})
	using the evolution of perturbations during matter domination
	\cite{Ali:2010gr}.
	Later,
	Creminelli {\etal} used an effective theory of this type
	to propose an alternative to the usual inflationary mechanism
	for generating a primordial density perturbation
	\cite{Creminelli:2010ba}.

	Kobayashi {\etal} proposed a kinetically-driven model of
	inflation supported by a Galileon-like field, which they called
	``G-inflation.''
	In this model, $\LagM$ was taken to be of the form
	$\LagM = P(X, \phi) + F(X,\phi) \Box \phi$, where
	$X = - \grad_\mu \phi \grad^\mu \phi$
	\cite{Kobayashi:2010cm}.
	Kobayashi {\etal} gave background solutions and studied the
	two-point correlation function of perturbations;
	three-point correlations in this model were later considered by
	Mizuno \& Koyama \cite{Mizuno:2010ag}.
	As we will explain,
	although this scenario is qualitatively similar to a Galileon model,
	the
	``G-inflation'' picture described
	in Ref.~\cite{Kobayashi:2010cm} relies on
	a hard breaking of the Galilean symmetry.
	This symmetry is essential to stabilize the theory against quantum
	corrections;
	if it is broken, there is
	no protection against the radiative generation of higher operators.
	These not only spoil the predictions of the theory,
	but lead to ghost-like instabilities. 

	In this paper we take a different approach.
	We argue that it is important to retain the Galilean symmetry,
	as far as possible.
	It will turn out that a non-trivial
	background geometry
	introduces soft breaking terms,
	which lead to an interesting phenomenology.
	However,
	Galileon models which preserve
	sufficient symmetry
	represent an
	alternative to the DBI Lagrangian
	as a ghost-free, radiatively stable
	higher-derivative inflationary model.
	In comparison with the studies of
	Kobayashi {\etal} and Mizuno \& Koyama
	we work with the most general covariant completion of the
	Galileon-invariant Lagrangian.

	We refer to an epoch of inflation driven by
	a Galilean-invariant field
	(perhaps broken by a nontrivial background)
	as `Galileon inflation.'
	We calculate the bispectrum nongaussianity parameter
	$\fnl$
	\cite{Komatsu:2001rj}
	and show that Galileon models can have an observational
	signature
	which is distinct from both DBI and canonical inflation.
	Indeed, Galileon models occupy a previously unexplored
	regime in the effective field theory
	of inflationary perturbations,
	where observable nonlinearities may be much larger
	than those obtained from even the highly nonlinear DBI Lagrangian.
	This difference arises because the cubic interactions are
	described by a combination of dimension-six \emph{and}
	dimension-seven operators, rather than the dimension-six operator
	alone, which would generically dominate.

	In what follows we work with signature $(-,+,+,+)$
	and allow Greek indices $\mu$, $\nu$, \ldots,
	to range over spacetime indices,
	whereas Latin indices
	$i$, $j$, \ldots, label purely spatial indices.
	We choose units in which the Dirac constant, $\hbar$,
	and the speed of light, $c$, are set to unity,
	and express the gravitational coupling in terms of the
	reduced Planck mass, $\Mp^2 = (8 \pi G)^{-1/2}$.
	Spacetime indices in complementary upper and lower positions are
	contracted with the spacetime metric, and $\Box = \grad^\mu \grad_\mu$.
	We reserve $\partial$ to denote a derivative with respect to
	purely spatial indices.
	Spatial indices are contracted using a Kronecker delta,
	so that $\partial^2 = \partial_i \partial_i$.
	
	\section{Inflation}
	\label{sec:inflation}
	
	In this section we briefly review the conditions
	$\LagM$ must satisfy to constitute a well-defined, predictive
	description of an inflationary phase.
	
	\para{Unitarity}
	An arbitrary action of the form~\eqref{eq:action}
	contains derivatives of second-order and higher applied to
	the elementary fields, and therefore leads to
	equations of motion of third-order or higher.
	Unless an infinite number of derivatives are present
	such theories do not usually possess a well-defined Cauchy problem
	\cite{Barnaby:2007ve},
	which manifests itself in the appearance of ghost states.
	At the quantum level the result is a pernicious loss of unitarity.
	Accordingly we should
	restrict our attention to special choices of $\LagM$ which lead
	to second-order equations of motion
	\cite{*[{Even where $\LagM$ satisfies this condition, coupling
	to gravity typically makes perturbations
	around the background interact through higher derivatives,
	or even inverse derivatives in certain gauges.
	Nevertheless, the propagating degrees of freedom are determined by
	the leading relevant operators, which comprise the
	quadratic operators $\phi^2$ and $(\grad \phi)^2$ if the
	expansion is around a Gaussian free theory.
	The modes propagated by these quadratic operators
	are stable and lead to a well-defined Cauchy
	problem, even if their interactions involve derivatives. }]
	[{}] Donoghue:1993eb, *Donoghue:1994dn,
	*[{}] [{. In our case this approach is not sufficient.
	We wish to work with theories where the leading relevant operators
	themselves involve higher derivatives. For such models there is
	no prescription which truncates the theory to a finite set
	of stably propagating excitations, and so the restriction
	to second-order equations of motion is mandatory.}]Weinberg:2008hq}.
	
	\para{Quantum fluctuations}
	It has recently been argued that the choice $\LagM = F(X,\phi) \Box \phi$
	(used in Ref.~\cite{Kobayashi:2010cm})
	leads to second-order equations of motion for any choice of $F$
	\cite{Deffayet:2010qz},
	and therefore
	unitary evolution as a quantum field theory.
	For this reason, one could contemplate such models as
	candidates for an inflationary action of the form~\eqref{eq:action}.
	
	However, unitarity is not the only condition for
	predictivity.
	A typical $F$ can be represented as a power series in $X$
	and $\phi$, with some specified coefficients. But we should recognize
	that in the absence of other symmetries there is no way
	to protect these coefficients from quantum corrections.
	An important example is the DBI
	Lagrangian $\LagM \supseteq f(\phi) \sqrt{1 - X}$
	for an arbitrary $f(\phi)$.
	In the limit $X \sim 1$, fluctuations around the background
	solution have small sound speeds $c_s \ll 1$
	and acquire large nongaussianities of order $\fnl \sim c_s^{-2}$
	\cite{Alishahiha:2004eh, Chen:2006nt}.
	But at $X \sim 1$ the square root formally receives contributions
	from all powers of $X$ with coefficients in precisely defined ratios,
	which in principle are susceptible to disruption by renormalization.
	The prediction $\fnl \sim c_s^{-2}$ is believed to be trustworthy
	only because a symmetry forbids
	large renormalizations of these coefficients
	\cite{Tseytlin:1999dj, Silverstein:2003hf}, forcing
	quantum corrections to involve
	the \emph{two}-derivative combination $\grad \grad \phi$.
	A similar symmetry protects fluctuations around
	models that exhibit Galilean symmetries
	(to be defined below), such as those that arise in
	the decoupling limit of either massive gravity or
	the Dvali--Gabadadze--Porrati (DGP) model,
	and can even be extended to theories containing multiple fields
	\cite{Dvali:2000hr,
	*Luty:2003vm, Porrati:2004yi, *Nicolis:2004qq, *Endlich:2010zj,
	Nicolis:2008in, deRham:2010eu,
	Hinterbichler:2010xn, Deffayet:2010zh, *Padilla:2010de,
	*Padilla:2010ir, *Padilla:2010dr, *deRham:2007rw, *deRham:2007xp}.
	We discuss these
	non-renormalization properties in \S\ref{sec:break}.
	
	The lesson we wish to draw from these examples is that
	our calculations are likely to be trustworthy only if
	the functional form of $\LagM$ is protected from
	large renormalizations by the presence of a symmetry.
	In seeking to generalize the DBI and Galileon-type Lagrangians, there
	are essentially only two choices.
	
	\begin{enumerate}
		\item The shift symmetry $\phi \rightarrow \phi + c$ is the
		only symmetry of the action.
		In this case we must take the leading relevant operators to be the
		quadratic mass and kinetic terms, $\phi^2$ and $(\grad \phi)^2$,
		and terms with more derivatives should be treated as irrelevant
		in the technical sense.
		Corrections to the Gaussian theory arise from at most a few
		non-renormalizable operators, leading to very small
		observable departures from purely Gaussian fluctuations
		\cite{Creminelli:2003iq, Seery:2005wm}.
		\label{choice:canonical}
		
		\item The shift symmetry is combined with a second symmetry,
		which protects the form of any derivative interactions.
		Such a symmetry cannot commute with the generators of the
		Poincar\'{e} group, because it necessarily mixes derivatives
		of different orders. Therefore it must be nonlinearly realized
		as a nonfactorizable extension of the Poincar\'{e} group.
		\label{choice:noncanonical}
	\end{enumerate}
	
	Choice \ref{choice:canonical} leads to canonical slow-roll inflation
	with very small corrections, and is well-understood.
	Interesting generalizations must therefore make use of
	\ref{choice:noncanonical} and can be classified according to their choice
	of nonfactorizable extension. Only a handful of such
	extensions are known, of which
	one is the conformal group. Others can be constructed
	beginning with the
	Poincar\'{e}, de Sitter or anti-de Sitter groups in dimension
	greater than four and applying Wigner--In\"{o}n\"{u}
	contractions.
	Following this reasoning one arrives at a short list
	of possible $\LagM$ which are protected by symmetries.
	The choices are
	DBI, warped-DBI, Galileon and conformal-Galileon models,
	together with their higher-dimensional
	generalizations
	\cite{Nicolis:2008in, deRham:2010eu}
		\footnote{This list is not necessarily exhaustive. Although
		these are the only known cases, other possibilities may exist.}.
	In this paper we focus on the general class of (conformal-) Galileon
	models. In such models the protecting
	`Galilean' symmetry is an extension of the
	shift symmetry to include spacetime translations,
	$\phi \rightarrow \deltaG \phi = \phi + b_\mu x^\mu + c$,
	where $b_\mu$ and $c$ are constant.
			
	\para{Observables}
	Eq.~\eqref{eq:action} implies
	a large degeneracy among
	$\LagM$ which realize given values of $\epsilon$
	and $\eta$, and therefore the
	amplitude $\Ps^{1/2}$ and spectral index $n_s$ of the
	primordial density perturbation \cite{Lyth:2009zz}.
	Over the last decade,
	significant effort has been invested
	in developing observables
	which distinguish one choice of $\LagM$ from
	another
	\cite{ArmendarizPicon:1999rj, *Garriga:1999vw, *ArkaniHamed:2003uz,
	Creminelli:2003iq, Seery:2005wm, Chen:2006nt}.
	As we now explain, the most powerful family of such observables is the
	amplitude of three-, four- and higher $n$-point correlations.
	
	How are inflationary models to be distinguished?
	The renormalization programme has taught us that predictions
	extracted from quantum field theories, such as those governing
	inflation,
	express measurable quantities in terms of a finite number
	of experimental inputs---%
	sometimes expressed as
	``observables in terms of observables.''
	In canonical inflation there are two relevant operators,
	$(\grad \phi)^2$ and $\phi^2$,
	which each require a parameter to be extracted from experiment.
	Typically these are
	$(H / \Mp)^2 / \epsilon$ and $\eta$,
	extracted from $\Ps^{1/2}$ and $n_s$.
	The ratio $(H /\Mp)^2$ alone can be determined from
	the tensor--scalar ratio $r$ and fixes a particular de Sitter background
	geometry.
	In a noncanonical model of the form~\eqref{eq:action},
	the action for fluctuations may exhibit spontaneous breakdown of
	Lorentz invariance,
	giving $\dot{\phi}^2$ and $(\partial \phi)^2$
	independent coefficients. This breaking is measured by
	the sound speed $c_s^2$.
	The gravitational coupling, $\Mp$,
	is measured by terrestrial gravitational experiments.
	
	We conclude that no measurement involving $\Ps^{1/2}$,
	$n_s$, and $r$ alone can tell us about the operators present in
	$\LagM$, although $c_s$ can
	diagnose whether $\dot{\phi}^2$ and
	$(\partial \phi)^2$ have
	independent coefficients.
	To distinguish one choice of $\LagM$ from another, we must ask
	whether further observables such as the bispectrum nongaussianity,
	$\fnl$, and the trispectrum observables $\tnl$ and $\gnl$
	\cite{Okamoto:2002ik, *Boubekeur:2005fj, *Sasaki:2006kq}
	can be expressed in terms of the input parameter set.
	Maldacena showed that, in a single-field model, this is true
	for $\fnl$ \cite{Maldacena:2002vr}
	unless an irrelevant operator associated with
	some higher slow-roll parameter is unexpectedly large.
	The same is true for the single-field trispectrum observables
	\cite{Seery:2006vu, *Seery:2008ax}.
	It is for this reason that nongaussianities are such a stringent
	test of the single-field framework, because measurements of
	$\fnl$, $\tnl$ and $\gnl$ cannot simply be absorbed into a new parameter
	of the model.

	Similar reasoning applies to noncanonical models described
	by~\eqref{eq:action}. There is a simplification
	in theories which are not approximately
	canonical inflation,
	to be discussed in \S\ref{sec:eft} below,
	because the influence of gravity becomes negligible
	\cite{Alishahiha:2004eh, ArkaniHamed:2003uz, Cheung:2007st,
	Senatore:2009gt, Bartolo:2010bj, *Bartolo:2010di}.
	Therefore observables involve the parameters
	of $\LagM$ alone, and do not mix with $\Mp$.
	Different choices of $\LagM$ may require a larger or smaller
	set of
	input parameters
	to express their predictions. However,
	once these are measured, the remaining observables
	are uniquely determined. Therefore
	these express genuine differences between one
	$\LagM$ and another.
	For example, in DBI inflation there is a firm prediction for
	$\fnl$ once $c_s$ has been determined, perhaps using the tilt of the
	tensor power spectrum \cite{Lidsey:2006ia}.
	We will show in \S\ref{sec:pdp} below that there is a different
	prediction for $\fnl$ in Galileon models, making
	the nongaussianity of the primordial
	density perturbation a particularly powerful observational
	probe of Galileon inflation.
	The situation is less favourable if the form of $\LagM$
	is not protected by a symmetry as with $\LagM \supseteq F(X,\phi)
	\Box \phi$ for general $F$.
	In that case, an infinite number of experimental inputs
	are required to determine $F$ and the theory loses predictivity.
	
	\para{Experimental limits}
	Measurements of the basic input parameters
	$\Ps^{1/2}$ and $n_s$ are now quite precise
	\cite{Komatsu:2010fb}, and will improve further once data arrive
	from the \emph{Planck} satellite.
	Limits on the bispectrum vary depending on the momentum
	dependence, or `shape'
	\cite{Babich:2004gb}.
	Present-day limits are moderately constraining
	\cite{Smith:2009jr, Komatsu:2010fb, Senatore:2009gt},
	and data from the Planck satellite is expected to furnish
	very stringent constraints. For this reason it is realistic
	to anticipate the role of
	$\fnl$ as a constraint on inflationary models.
	Limits on the trispectrum parameters are presently weak
	\cite{Smidt:2010ra}.
	Smidt {\etal} estimate that an optimistic future experiment may constrain
	$\fnl$ to $\pm 3$, $\tnl$ to $\pm 225$ and $\gnl$ to $\pm 6 \times 10^4$.
	It is not yet clear whether the amplitude of higher correlations
	can ever be significantly constrained, so these parameters
	are likely to exhaust the available input parameters for at least
	the near future.
		
	If $\LagM$ requires more input parameters than can be measured,
	the model again becomes unpredictive.
	This contrasts with accelerator
	experiments where---in principle---there
	is typically no limit to the number of parameters
	which can be taken from experiment. In this paper we wish to
	emphasize the distinction of \emph{principle} between a predictive theory,
	where a symmetry picks out a finite number of
	invariant operators whose coefficients fix the other observables,
	and unpredictive theories where quantum effects imply that an infinite
	number of coefficients must be retained. In practice, however,
	a failure of predictivity for whatever cause is problematic.
	
	\section{Breaking the shift symmetry}
	\label{sec:break}
	
	An exact shift symmetry in Eq.~\eqref{eq:action} makes the
	value of $\phi$ unphysical. To proceed we softly break the symmetry
	by hand, giving $\phi$ a dynamical vacuum expectation value.
	We first argue that this breaking is necessary because we would
	otherwise obtain $\dot{\phi} = 0$, preventing the existence of
	a conserved curvature perturbation $\zeta$ at long wavelengths.
	
	The mildest possible breaking is to introduce a potential
	$V = V_0 - \lambda^3 \phi$, where $V_0$ is a constant and
	$\lambda$ has engineering dimension $[\text{mass}]$.
	Taking the decoupling limit $\Mp \rightarrow \infty$ while keeping
	the de Sitter background fixed,
	so that $3 H^2 \Mp^2 = V_0$,
	we recover a scalar field theory on exact de Sitter space.
	In this limit, a term linear in $\phi$ does not break the shift
	symmetry because the action transforms as a total derivative,
	\begin{equation}
		\delta_c \left[
			\sqrt{-g} \; \phi
		\right]
		= a(t)^3 c = \frac{1}{3} \partial_i
		\left[ a(t)^3 c\, x_i \right] .
	\end{equation}
	We conclude that with dynamical gravity the symmetry must be broken
	softly in the sense that its effects are suppressed by powers of
	$1/\Mp$.
	
	Because the shift symmetry is exact in the decoupling limit,
	we conclude that the equation of motion for $\phi$
	can be written as a current conservation equation
		\footnote{A similar observation was made in the recent preprint
		Ref.~\cite{Hinterbichler:2010xn}, which appeared as this paper
		was in the final stages of preparation.}
	\begin{equation}
		\grad_\mu J^\mu = \dot{J}^t + 3 H J^t = \lambda^3 ,
		\label{eq:Jconserve}
	\end{equation}
	where $J^\mu$ is the Noether current,
	\begin{equation}
		J^\mu = \frac{\partial \LagM}{\partial ( \grad_\mu \phi )}
		- \grad_\nu \left[
			\frac{\partial \LagM}{\partial ( \grad_\nu \grad_\mu \phi )}
		\right]
		+ \cdots .
		\label{eq:Jdef}
	\end{equation}
	Eq.~\eqref{eq:Jconserve} admits the exact solution $J^t = \lambda^3 / 3H$,
	which could equally have been obtained by passing to the overdamped
	limit $\dot{J}^t / J^t \rightarrow 0$.
	Since $J^t$ is constant, the shift symmetry allows us to infer that
	$\dot{\phi}$ is also constant
	and proportional to $\lambda^3$.
	(To reach this conclusion we require the
	theory to have a well-posed initial value formulation, which is
	guaranteed by our restriction to second-order equations of motion.)
	In the unbroken case where $\lambda = 0$ we would obtain
	$\dot{\phi} = 0$.

	At first, it
	might seem surprising that inflation in this scenario
	is possible
	only in the presence of a potential $\lambda \ne 0$.
	It is now well-understood that models of
	``$k$-inflation'' type
	\cite{ArmendarizPicon:1999rj, *Garriga:1999vw}
	can achieve an accelerated regime in purely kinetic scenarios,
	where
	$\LagM = \LagM(X)$.
	The resolution is that the equations of motion admit
	solutions of two types: a transient decaying contribution---which,
	in perturbation theory,
	is the decaying mode; and a dominant growing-mode solution, which is the
	one we consider above.
	Purely kinetic $k$-inflation models
	make use of the normally transient
	decaying mode to source the background
	expansion. In what follows we shall work with the more conventional
	solution.

	\para{Nonrenormalization of mass}
	The necessity of including a nontrivial potential is problematic.
	Since potential terms are not invariant under the shift symmetry,
	they must be treated as irrelevant operators which
	induce small corrections. But operators of
	mass dimension less than four would typically receive large
	quantum corrections and, if present, such large renormalizations
	destroy the radiative stability of $\LagM$,
	making the model of limited interest.
	In practice we cannot avoid this issue because
	a potential term of at least quadratic
	order is required
	to describe a graceful exit from the inflationary phase.

	Remarkably, the operators $\phi$ and $\phi^2$ are protected by
	a nonrenormalization theorem and therefore
	can be treated consistently as irrelevant deformations of the
	shift-invariant Lagrangian
	\cite{Porrati:2004yi, *Nicolis:2004qq, *Endlich:2010zj}.
	This conclusion may be reached most straightforwardly
	by considering the
	one-particle irreducible effective action, $\Gamma[\phi]$,
	\begin{equation}
		\exp{\im \Gamma[\phi]}= \int \D\psi \;
		\exp
			\left\{
				\im S[\phi+\psi]
				- \im \psi \frac{\delta \Gamma[\phi]}{\delta \phi}
			\right\} .
	\end{equation}
	We separate $\Gamma[\phi]$ into its classical part, which
	coincides with the classical action $S[\phi]$,
	and a quantum part $\Gammaloop$, writing
	$\Gamma[\phi]=S[\phi]+ \Gammaloop[\phi]$. Therefore
	\begin{equation}
	\begin{split}
		\exp \im \Gammaloop[\phi]
		=
		\int \D\psi \;
		\exp \Bigg\{ &
			\im S[\phi+\psi]
			- \im S[\phi]
		\\ & \mbox{}
			- \im \psi \frac{\delta S[\phi]}{\delta \phi}
			- \im \psi \frac{\delta \Gammaloop[\phi]}{\delta \phi}
		\Bigg\} .
	\end{split}
	\end{equation}
	First,
	consider an action of the form
	\begin{equation}
		S[\phi]= S_0[\phi] + \int \d^4 x \;
			\left( \lambda \phi + \frac{1}{2}m^2 \phi^2 \right) .
		\label{eq:protected}
	\end{equation}
	We assume that $S_0[\phi]$ is invariant under the
	Galilean symmetry, so that $S_0[\deltaG \phi]=S_0[\phi]$,
	where $\deltaG \phi = \phi + b_\mu x^\mu + c$ is the
	Galileon shift operator defined in \S\ref{sec:inflation}.
	It follows that
	\begin{equation}
	\begin{split}
		\exp \mbox{} & \im \Gammaloop[\phi] =
		\int \D\psi \;
		\exp
			\im \Bigg\{
				S_0[\phi+\psi]
				- S_0[\phi]
		\\ & \mbox{}
		+ \int \d^4 x \; \frac{1}{2}m^2 \psi^2
		- \psi \frac{\delta S_0[\phi]}{\delta \phi}
		- \psi \frac{\delta \Gammaloop[\phi]}{\delta \phi}
		\Bigg\} .
	\end{split}
	\end{equation}
	Therefore we
	conclude $\Gammaloop[\deltaG \phi]=\Gammaloop[\phi]$.
	Hence, even though a mass term \emph{explicitly}
	breaks the Galileon symmetry, this does not induce any further operators
	which violate the symmetry at the quantum level. 	
	
	The same conclusion can be reached by analysing the properties of
	Feynman diagrams. Consider any nontrivial diagram with two external lines,
	which in principle could contribute to a renormalization of $\phi^2$.
	Such a diagram includes two field operators where the external
	lines attach to the interior of the diagram,
	and the structure of $\LagM$ implies that each of these operators carries
	at least \emph{one} derivative.
	We conclude that the net effect of these field operators
	must introduce an overall factor of
	at least two powers of the external momenta.
	After expanding into a series of operator products,
	each term must contain at least \emph{two} derivatives
	and cannot include the operator $\phi^2$, contrary to our original
	supposition. As a trivial special case we recover the obvious fact
	that a $\phi^2$ operator, considered as a deformation of the Gaussian
	kinetic term $\LagM = X/2$, is not renormalized.
	
	The same argument does not apply
	if we extend~\eqref{eq:protected} to include higher powers of $\phi$.
	In particular, inclusion of a $\phi^3$ operator
	or higher would typically renormalize the coefficient of $\phi^2$ to
	$\sim \Lambda^2$, where $\Lambda$ is the
	cutoff of the theory. In such circumstances we would be obliged to take
	$\phi^2$ as a relevant operator, leading to strong radiative breaking
	of the supposed shift symmetry exhibited by $\LagM$.
	Therefore inclusion of such
	operators is inconsistent.
	
	The conclusion of these arguments is that it is possible to construct an
	inflationary model based on a Galileon field, in which the shift and
 	Galilean symmetry protects the form of the Lagrangian.
 	Inflation can end, despite the presence of these symmetries,
 	at least in the decoupling limit $\Mp \rightarrow \infty$.
 	Although the Galileon symmetry is broken when coupled to gravity,
 	the breaking terms will be parameterically suppressed by powers of
 	$\Lambda/\Mp$, which we assume to be small.
	In what follows we will allow a little extra freedom and work
	with an arbitrary potential $V(\phi)$.
	Although this will generally break the Galilean symmetry explicitly,
	the foregoing argument demonstrates that any operators
	generated in this way will be suppressed by powers of three or more
	derivatives of $V(\phi)$. We will suppose that the models used to
	obtain Galileon inflation all break the Galilean symmetry mildly in
	this above sense.	
	
	\section{Galileon inflation}
	\label{sec:gal-inflation}

	On a curved background, such as de Sitter space,
	Deffayet {\etal}
	\cite{Deffayet:2009wt, *Deffayet:2009mn}
	remarked that
	the Galileon action
	constructed by Nicolis {\etal} \cite{Nicolis:2008in}
	leads to unwanted higher-derivative equations of motion,
	spoiling the expected construction of a ghost-free, unitary
	theory. This can be cured using a nonminimal coupling to gravity,
	which Deffayet {\etal} described as `covariantization.'
	The covariant Galileon action,
	which can also be
	obtained from the five-dimensional covering theory
	\cite{deRham:2010eu},
	is
	\begin{widetext}
		\begin{equation}
		\begin{split}
			S = & \int \d^4 x \; \sqrt{-g} \, \Bigg[
				- \frac{c_2}{2} (\grad \phi)^2
				+ \frac{c_3}{\Lambda^3} \Box \phi ( \grad \phi )^2
				- \frac{c_4}{\Lambda^6} ( \grad \phi )^2 \Big\{
					(\Box \phi)^2 - (\grad_\mu \grad_\nu \phi)
					(\grad^\mu \grad^\nu \phi)
					- \frac{1}{4} R (\grad \phi)^2
				\Big\}
			\\ & \mbox{}
				+ \frac{c_5}{\Lambda^9} ( \grad \phi )^2 \Big\{
					(\Box \phi)^3 - 3 (\Box \phi)( \grad_\mu \grad_\nu \phi)
					(\grad^\mu	 \grad^\nu \phi)
					+ 2 ( \grad_\mu  \grad_\nu \phi)
					(\grad^\nu	 \grad^\alpha \phi)
					(\grad_\alpha \grad^\mu \phi)
					- 6 G_{\mu \nu} \grad^\mu \grad^\alpha \phi
					\grad^\nu \phi \grad_\alpha \phi
				\Big\}
			\Bigg] ,
		\end{split}
		\label{eq:galileon}
		\end{equation}
	\end{widetext}
	where and $G_{\mu\nu}$ and $R$ are respectively the Einstein tensor and
	scalar curvature of the background.
	Typically, both
	covariantization and inclusion of a non-Minkowski background metric
	will
	softly break the Galilean symmetry.
	It will emerge that this has important
	consequences for the phenomenology of the model.
	The coefficients $c_i$ are dimensionless, and---as above---$\Lambda$ is a
	mass scale which
	determines the na\"{\i}ve cutoff of the theory.
	In practice, fluctuations around a nontrivial background
	can be valid up to energies somewhat larger than
	$\Lambda$ if a Vainshtein effect is operative
	(see Ref.~\cite{Vainshtein:1972sx}),
	discussed in more detail in Ref.~\cite{deRham:2010eu}.
	Had we included nonminimal coupling to the geometry in
	Eq.~\eqref{eq:action}, the curvature terms involving $G_{\mu\nu}$
	and $R$ would have been accompanied by other geometric invariants,
	but all such terms are suppressed by powers of $H/\Lambda$
	(\emph{cf}. Eqs.~\eqref{eq:alpha}--\eqref{eq:ffour}, where
	such suppressed contributions can be clearly identified).
	In the inflationary regime of interest, where nonlinearities
	are dominated by the Galileon self-interactions, it will transpire that such
	terms are negligible, but for completeness we continue to retain
	the nonminmal curvature couplings required by covariantization.

	The action for fluctuations in the decoupling limit of the
	DGP model has $c_4 = c_5 = 0$.
	Constraints on the $c_i$ obtained from short-distance gravitational
	effects
	were studied in Refs.~\cite{Nicolis:2008in, Burrage:2010rs}
	for the case that Eq.~\eqref{eq:galileon} describes the short-distance
	effects of a modification of gravity today.
	If the scalar $\phi$ is taken to be relevant only during inflation,
	however, the $c_i$ are unrestricted and must be determined
	independently from cosmological probes.
	Cosmological constraints on $c_2$, $c_3$ and $c_4$ are quoted by
	Ali {\etal} \cite{Ali:2010gr}.

	\subsection{Infrared completion of the Galileon}

	Models of this type are unusual, because they must be viewed as
	effective field theories both in the usual sense of being valid
	below some energy scale $\Lambda$, and
	also in the sense that they may require a non-trivial
	infrared completion.
	For instance, as described in \S\ref{sec:intro},
	they typically arise as intermediate-scale effective theories
	corresponding to massive gravities, and are therefore
	modified at an infrared scale set by the Compton wavelength
	of the graviton.
	Such models were discussed in
	Refs.~\cite{deRham:2010eu, deRham:2010gu, *deRham:2010ik}.
	The consistency of our present analysis only requires that
	the infrared cutoff is much larger than the Hubble scale during
	inflation. However,
	in a full cosmological model accounting
	for the subsequent expansion of the universe,
	it is likely that
	this scale must be many orders of magnitude larger.
	We do not address this issue in this paper.

	\subsection{Inflation in the de Sitter decoupling limit}
	\label{sec:DGP}

	To get a sense of the background solutions and new features
	in Galileon models, consider the background field evolution
	in a de Sitter decoupling limit, defined by the limit
	$\Mp \rightarrow \infty$ where $H$ is kept fixed.
	(We caution that
	this should not be confused with the decoupling limit considered in the
	next section, which allows for a more general FRW background).
	This limit is applicable if the variation $\Delta V$ in the inflationary
	potential over the duration of inflation satisfies $|\Delta V/V| \ll 1$.
	In this limit we have a Galileon model living on the background
	of de Sitter spacetime, with scale factor $a(t)=\e{Ht}$.

	After integrating by parts to obtain the action in first-order
	form, and cancelling any boundary terms generated by this process
	\cite{Dyer:2009yg},
	the action for
	a homogeneous field configuration $\phi(t)$ can be written
	\begin{equation}
	\begin{split}
		S_0 = \int \d^4 x \; a^3 \Bigg\{ &
			\frac{c_2}{2} \dot{\phi}^2
			+ \frac{2c_3 H}{\Lambda^3} \dot{\phi}^3
			+\frac{9c_4H^2}{2\Lambda^6} \dot{\phi}^4
		\\ & \mbox{}
			+\frac{6c_5H^3}{\Lambda^9} \dot{\phi}^5
			+ \lambda^3 \phi
		\Bigg\} .
	\end{split}
	\label{eq:ds-action}
	\end{equation}
	According to
	the discussion of Eqs.~\eqref{eq:Jconserve}--\eqref{eq:Jdef},
	the current $J^t$ can be written
	\begin{equation}
		J^t = \dot{\phi}
			\left(
				c_2
				+ 6 c_3 Z
				+ 18 c_4 Z^2
				+ 30 c_5 Z^3
			\right)
		=
			\frac{\lambda^3}{3H} .
		\label{eq:z-current}
	\end{equation}
	where $Z$ is the dimensionless combination
	\begin{equation}
		Z \equiv \frac{H \dot{\phi}}{\Lambda^3} .
		\label{eq:z-def}
	\end{equation}
	We shall see that $Z$ plays the role of coupling constant in
	Galileon theories. 

	There are two regimes. First, a weakly coupled solution
	for which $Z \ll 1$,
	\begin{equation}
		\dot{\phi} \sim \frac{\lambda^3}{3c_2 H}
		\label{eq:weak}
	\end{equation}
	In this regime the outcome is very close to canonical slow-roll
	inflation.
	Second, there is a strongly coupled regime
	for which $Z \gg 1$.
	Specializing to a DGP-like Galileon theory
	\cite{Dvali:2000hr}
	for which only $c_2$ and $c_3$ are nonzero,
	the strongly coupled regime gives
	\begin{equation}
		\dot{\phi} \sim \sqrt{ \frac{\Lambda^3 \lambda^3}{18c_3 H^2} } .
		\label{eq:strong}
	\end{equation}
	The smooth solution interpolating between Eqs.~\eqref{eq:weak}
	and~\eqref{eq:strong} is
	\begin{equation}
		\dot{\phi} = \frac{\Lambda^3}{12H} \frac{c_2}{c_3}
			\left( -1 + \sqrt{1 + 8 \frac{c_3}{c_2^2}
				\frac{\lambda^3}{\Lambda^3}}
			\right) .
	\end{equation}
	More generally, when the other Galilean interactions are present,
	the field configuration
	interpolates between the weak coupling regime \eqref{eq:weak} and the
	strongly coupled
	solution $\dot \phi H\sim(\lambda \Lambda^2)$ if $c_4\ne 0$ or
	$\dot \phi H\sim(\lambda^3 \Lambda^9)^{1/4}$ if $c_5\ne 0$.
	
	What is the relevance of corrections to~\eqref{eq:ds-action}
	from nonminimal coupling to the geometry,
	which were briefly discussed below Eq.~\eqref{eq:galileon}?
	To be concrete we consider a coupling of the form $\phi G$,
	where $G$ is the Gauss-Bonnet invariant,
	which would retain the important feature of second-order
	equations of motion.
	Since $G$ involves the square of contractions of the Riemann tensor,
	the leading contribution must be of the form
	$\sim H^3 \dot{\phi} / \Lambda$.
	This will be of comparable importance to
	the $c_3$-dependent term $\sim H \dot{\phi}^3 / \Lambda^3$
	only if
	\begin{equation}
		\frac{1}{Z^2} \frac{H^4}{\Lambda^4} \gtrsim 1 .
		\label{eq:gb-importance}
	\end{equation}
	Unless $Z < (H/\Lambda)^2 \ll 1$,
	Eq.~\eqref{eq:gb-importance}
	and similar conditions for the $c_4$ and $c_5$-terms
	show that the Gauss--Bonnet invariant is negligible compared to the
	terms proportional to $c_i$, arising from Galileon operators.
	On the other hand, the relative importance of the higher-derivative
	Galileon operators to the lower-derivative operators is controlled by
	positive powers of $Z$.

	This has a simple interpretation.
	When $Z \gtrsim 1$ the nonlinearities of the Galileon sector
	are important, as in Eqs.~\eqref{eq:z-current} and~\eqref{eq:strong}.
	In this limit the mixing with gravity can be neglected,
	and non-minimal couplings such as $\phi G$ are irrelevant.
	It is in this regime that an interesting Vainshtein effect
	can emerge.
	In addition, as
	we will explain below, the inflationary phenomenology is rather different to
	canonical slow-roll models.
	On the other hand,
	the limit $Z \ll 1$, as in Eq.~\eqref{eq:weak}, describes weakly coupled
	perturbations in the Galileon sector,
	giving a theory almost equivalent to canonical
	slow-roll inflation.
	In this limit the strongest interactions come from mixing with
	gravity, and a coupling such as $\phi G$ cannot be neglected.

	\subsection{Effective field theory for inflation}
	\label{sec:eft}

	How does an era of Galileon inflation
	differ from a canonical inflationary phase?
	We have argued in \S\ref{sec:inflation} that two Lagrangians are
	inequivalent only if they make different predictions for observables.
	Therefore we must study perturbations generated
	by the action~\eqref{eq:action}, which are the
	appropriate measurable quantities.
	
	\para{Unitary gauge action}
	Cheung {\etal} \cite{Cheung:2007st}
	argued that certain properties of the perturbations
	generated in an inflationary model were fixed by the background,
	and were therefore model independent,
	whereas others varied between theories and could be used to probe
	different choices of $\LagM$.
	This conclusion was obtained by constructing the most general
	action for small fluctuations on a quasi-de Sitter background,
	subject to the condition of unbroken spatial diffeomorphisms
	and nonlinearly realized Lorentz invariance.
	In \S\ref{sec:flucts} below we will construct the action for small
	fluctuations in Galileon inflation using a more direct approach.
	However, this action could equally
	have been obtained by specializing the result
	of Ref.~\cite{Cheung:2007st}
	to a scenario with Galilean symmetries. Therefore,
	before proceeding with a detailed calculation,
	it is of interest to determine what constraints are placed on the
	model by the construction of Cheung {\etal}
	
	The authors of Ref.~\cite{Cheung:2007st}
	worked in a model with a single scalar field,
	$\phi$, and constructed their action in a gauge where
	slices of constant time coincided with slices of uniform $\phi$.
	In this gauge there are no explicit
	scalar fluctuations, but only perturbations of the metric.
	The unit vector normal to slices of constant time is $n^\mu$,
	and constitutes a preferred vector field breaking manifest
	Lorentz invariance.
	The Lagrangian for inflationary perturbations can be built only out of
	operators which are invariant under spatial diffeomorphisims
	associated with reparametrizations of the induced three-dimensional
	spatial metric
	$h_{\mu\nu} = g_{\mu\nu} + n_\mu n_\nu$.
	Cheung {\etal} showed that it was sufficient to take the Lagrangian
	to comprise a general scalar combination of the Riemann tensor,
	${R^\mu}_{\nu\rho\sigma}$,
	together with the time--time component of the metric,
	$g^{00}$, and the extrinsic curvature,
	$K_{\mu\nu} = - {h_{(\mu}}^\rho
	{h_{\nu)}}^\sigma \grad_\rho n_\sigma$,
	associated with
	slices of constant time \cite{Cheung:2007st}.
	Bartolo {\etal} argued that the
	most general Lagrangian
	including terms of up to cubic order
	in small fluctuations can be written
	\cite{Bartolo:2010bj, *Bartolo:2010di}
	\begin{widetext}
	\begin{equation}
	\begin{split}
		S =
		\int \d^4x \; \sqrt{-g} \;
		\Bigg[ &\
			\frac{1}{2}\Mp^2 R
			- c(t) g^{00}
			- \Lambda(t)
			+ \frac{1}{2} M_2(t)^4 (g^{00}+1)^2
			+ \frac{1}{3} M_3(t)^4 (g^{00}+1)^3
		\\ & \mbox{}
		- \frac{\bar{M}_1(t)^3}{2} (g^{00}+1) \delta {K^\mu}_{\mu}
		- \frac{\bar{M}_2(t)^2}{2} ({\delta K^\mu}_\mu)^2
		- \frac{\bar{M}_3(t)^2}{2} \delta K^{\mu\nu} \delta K_{\mu\nu}
		\\ & \mbox{}
		- \frac{\bar{M}_4(t)^3}{2} (g^{00}+1)^2 \delta {K^\mu}_{\mu}
		- \frac{\bar{M}_5(t)^2}{2} (g^{00}+1) ({\delta K^\mu}_\mu)^2
		- \frac{\bar{M}_6(t)^2}{2} (g^{00}+1)
			\delta K^{\mu\nu} \delta K_{\mu\nu}
		\\ & \mbox{}
		- \frac{\bar{M}_7(t)}{2} ({\delta K^\mu}_\mu)^3
		- \frac{\bar{M}_8(t)}{2} ({\delta K^\mu}_\mu)
			(\delta K^{\rho \sigma} \delta K_{\rho\sigma})
		- \frac{\bar{M}_9(t)}{2} \delta K^{\mu \nu} \delta K_{\nu \sigma}
			{\delta K^\sigma}_\mu
		\Bigg] ,
	\end{split}
	\label{eq:unitaryL}
	\end{equation}
	\end{widetext}
	where we have used the Riemann tensor only in the form of the
	Ricci scalar, to match low-energy gravitational
	experiments which probe the Einstein action.
	
	Following Refs.~\cite{Cheung:2007st, Bartolo:2010bj, *Bartolo:2010di},
	in writing Eq.~\eqref{eq:unitaryL} we have organized the
	expansion in powers of
	$\delta g^{00} = g^{00}+1$ and $\delta K_{\mu\nu}$, which are
	fluctuations
	around an unperturbed Friedmann--Robertson--Walker geometry.
	Therefore only the operators multiplying $c(t)$ and $\Lambda(t)$
	are non-zero on the background, which fixes these coefficients in
	terms of the expansion history $H(t)$.
	The $M_i(t)$ and $\bar{M}_i(t)$ are not fixed by the background
	evolution and encode differences between
	models determined by our choice of
	$\LagM$. It is in the effects generated by these operators that
	we should look for distinctive
	signatures of Galileon inflation.
	
	To convert Eq.~\eqref{eq:unitaryL} into a form suitable for computation
	it is helpful to make a gauge transformation,
	$t \rightarrow \tilde{t} = t - \pi(\vect{x},t)$.
	After this transformation,
	the equal-time hypersurfaces of the
	uniform-field (`unitary') gauge are deformed by $\pi$,
	which
	Cheung {\etal} argued could be considered as the Goldstone
	boson associated with broken time translation invariance.
	Each term in Eq.~\eqref{eq:unitaryL} generates
	a series
	in
	powers of $\pi$, with each copy of $\pi$ carrying least
	one gradient in time or space.
	Each copy of $\delta K_{\mu\nu}$ may additionally contribute
	a single
	power of $\pi$ carrying two gradients.
	It follows that Eq.~\eqref{eq:unitaryL} generates cubic interactions
	involving three copies of
	$\pi$ with between three and six derivatives.
	In this way a description of the system is constructed in terms of
	the lowest dimension operators compatible with the underlying
	symmetries---the effective field theory approach.
	
	\para{Decoupling limit}
	Eq.~\eqref{eq:unitaryL} is quite generally applicable and accounts for
	the mixing between $\pi$ fluctuations and the metric.
	However,
	when large nonlinearities are associated with the self-interactions
	of $\pi$ there exists a decoupling limit in which reliable predictions
	can be extracted while neglecting gravity
	\cite{Alishahiha:2004eh, ArkaniHamed:2003uz, Cheung:2007st,
	Senatore:2009gt}.
	For this limit to be a reasonable approximation,
	the $\Mp$-suppressed terms which are neglected must be smaller than
	terms which do not vanish when $\Mp \rightarrow \infty$.
	
	We briefly recapitulate the argument of
	Ref.~\cite{Cheung:2007st}.
	The most relevant kinetic term for the metric fluctuation
	$\delta g^{00}$ will come from the Ricci scalar. Therefore we can
	pass to canonical normalization by the rescaling
	$\delta g^{00} \rightarrow \delta g^{00}_c = \Mp \delta g^{00}$.
	The most relevant kinetic term for $\pi$ will
	arise from some nonlinear operator in Eq.~\eqref{eq:unitaryL}.
	If it is minimal
	in derivatives it will be of the form
	$M^4 \dot{\pi}^2$, where $M$ is some combination of the scales $M_i$
	or $\bar{M}_i$.
	The canonically normalized field is $\pi_c = M^2 \pi$.
	At quadratic level, a mixing term such as $M^4 \dot{\pi} \delta g^{00}$
	is negligible in comparison with the $\pi$ kinetic term
	for wavenumbers $k$ which satisfy $k \gtrsim E_{\mathrm{mix}} = M^2 / \Mp$.
	The same applies for cubic terms, where the leading mixing term
	$M^4 \dot{\pi}^2 \delta g^{00}$ is negligible in comparison
	with $M^4 \dot{\pi}^3$ under the same condition.
	A similar argument can be given if the most relevant kinetic term
	for $\pi$ contains higher derivatives \cite{Cheung:2007st},
	or if the leading cubic terms enter with a mass scale different
	from $M$. In the decoupling limit the scale $E_{\mathrm{mix}}$ must be
	smaller than the Hubble scale during inflation, making our
	predictions accurate to a relative error of order
	$E_{\mathrm{mix}}/ H$.
	In what follows we work in this limit,
	in which
	the metric can be taken to be
	unperturbed.
	Therefore it is most convenient to work in the uniform curvature
	gauge, where the unperturbed metric is spatially flat and can be
	taken as the background de Sitter geometry.
	
	\subsection{Fluctuations}
	\label{sec:flucts}

	On the basis of the foregoing discussion,
	we should
	study the effect of fluctuations around a cosmological background
	solution in the decoupling limit
	by constructing small fluctuations $t \mapsto t + \xi(\vect{x},t)$
	on a hypersurface of constant time,
	working in a gauge where such hypersurfaces are spatially flat.
	Note that this gives $\xi$ an engineering dimension
	$[\text{mass}]^{-1}$.
	
	The comoving curvature perturbation, $\zeta$,
	satisfies $\zeta = H \xi$
	and will be conserved on superhorizon scales, where
	$k c_s / aH \rightarrow 0$.
	Working to cubic order in $\xi$, the action can be written
	\begin{equation}
	\begin{split}
		S \supseteq \int & \d^4 x \; a^3 \Bigg[
			\alpha \dot{\xi}^2 - \frac{\beta}{a^2} ( \partial \xi )^2
			+ f_1 \dot{\xi}^3
			\\ & \mbox{}
			+ \frac{f_2}{a^2} \dot{\xi}^2 \partial^2 \xi
			+ \frac{f_3}{a^2} \dot{\xi} (\partial \xi)^2
			+ \frac{f_4}{a^4} (\partial \xi)^2 \partial^2 \xi
		\Bigg] ,
	\end{split}
	\label{eq:gal-action}
	\end{equation}
	where the symbol `$\supseteq$' is used to denote that $S$ contains these
	contribution among other higher-order ones,
	and the time-dependent coefficients $\alpha$, $\beta$
	and $f_i$ satisfy
	\begin{align}
		\alpha & = \frac{\dot{\phi}^2}{2} \left(
			c_2 + 12 c_3 Z + 54 c_4 Z^2 + 120 c_5 Z^3
			\right) ,
		\label{eq:alpha}
		\\ \nonumber
		\beta & = \frac{\dot{\phi}^2}{2} \Bigg\{
			c_2 + 4 c_3 ( 2 Z + \frac{\ddot{\phi}}{\Lambda^3})
			\\ \nonumber & \quad \mbox{}
			+ 2 c_4 \left[
				13 Z^2 + \frac{6}{\Lambda^6} \left( \dot{H} \dot{\phi}^2 +
				2 H \dot{\phi} \ddot{\phi} \right)
			\right]
			\\ & \quad \mbox{}
			+ \frac{24 c_5}{\Lambda^9} H \dot{\phi}^2 \left[
				2 \dot{\phi}(H^2 + \dot{H}) + 3 H \ddot{\phi}
			\right] \Bigg\} ,
		\label{eq:beta}
		\\
		f_1 & = \frac{2 H \dot{\phi}^3}{\Lambda^3}
			\left( c_3 + 9 c_4 Z + 30 c_5 Z^2 \right) ,
		\label{eq:fone}
		\\
		f_2 & = - \frac{2 \dot{\phi}^3}{\Lambda^3}
			\left( c_3 + 6 c_4 Z + 18 c_5 Z^2 \right) ,
		\label{eq:ftwo}
		\\ \nonumber
		f_3 & = - \frac{2 H \dot{\phi}^3}{\Lambda^3}
			( c_3 + 7 c_4 Z + 18 c_5 Z^2 )
			\\ & \quad \mbox{}
			+ \frac{2 \dot{\phi}^2 \ddot{\phi}}{\Lambda^3}
			( c_3 + 6 c_4 Z + 18 c_5 Z^2 ) ,
		\label{eq:fthree}
		\\ \nonumber
		f_4 & = \frac{\dot{\phi}^3}{\Lambda^3} \left\{
			c_3 + 3 c_4 Z + 6 c_5 \left[
				Z^2 + \frac{\dot{H} \dot{\phi}^2}{\Lambda^6}
			\right] \right\}
			\\ & \quad \mbox{}
			+ \frac{3 \dot{\phi}^3 \ddot{\phi}}{\Lambda^6}
			(c_4 + 4 c_5 Z) .
		\label{eq:ffour}
	\end{align}
	The quantity $Z$ was defined in
	Eq.~\eqref{eq:z-def}.
	In order that Galileon self-couplings dominate the interactions,
	and mixing with gravity can be neglected, we must have have
	$Z \gtrsim 1$. Although one can contemplate the limit
	$Z \gg 1$, there is some risk that this would spoil inflation
	unless renormalized by a Vainshtein effect.
	In this paper we restrict our attention to the case
	$Z \sim 1$ where nonlinearities are significant but not
	problematic.

	The contribution proportional to $\dot{\xi}^2 \partial^2 \xi$ can be
	removed after a field redefinition. Making the transformation
	$\xi \rightarrow \pi = \xi + f_2 \dot{\xi}^2 / 2 \beta$,
	it follows that Eq.~\eqref{eq:gal-action} can be written
	\begin{equation}
	\begin{split}
		S \supseteq
		\int \d^4 x \; a^3 \Bigg[ &
			\alpha \left\{ \dot{\pi}^2 - \frac{c_s^2}{a^2} (\partial \pi)^2
			\right\}
			+ g_1 \dot{\pi}^3
			\\ & \mbox{}
			+ \frac{g_3}{a^2} \dot{\pi} ( \partial \pi )^2
			+ \frac{g_4}{a^4} ( \partial \pi )^2 \partial^2 \pi
		\Bigg] ,
	\end{split}
	\label{eq:pi-action}
	\end{equation}
	where $g_3 = f_3$, $g_4 = f_4$,
	we have defined $c_s^2 = \beta/\alpha$, and
	$g_1$ is defined by
	\begin{equation}
		g_1 = f_1 + \frac{2 H}{c_s^2} f_2
		+ \frac{2}{3 c_s^2} \frac{\dot{\alpha}}{\alpha} f_2
		- \frac{\beta}{3 c_s^2}
		\frac{\d}{\d t} \left( \frac{f_2}{\beta} \right) .
		\label{eq:gone-def}
	\end{equation}
	Had we worked from the uniform-field gauge action of Cheung
	{\etal}, Eq.~\eqref{eq:unitaryL}, we would have obtained
	$\alpha$, $c_s$ and the $g_i$ in terms of $c(t)$, $\Lambda(t)$
	and the theory-dependent scales $M_i(t)$ and $\bar{M}_i(t)$.
	
	Although $\xi$ and $\pi$ will differ on small scales,
	they become equal whenever $\dot{\xi} = 0$ and are therefore equal
	in any epoch when $\xi$ is conserved. In particular, they coincide
	on superhorizon scales.
	Therefore, to obtain the correlation functions of
	$\zeta$ it suffices to obtain the correlation functions of $\pi$.

	\para{Relation to EFT action}
	The Galileon Lagrangian, constructed above,
	could have been recovered from
	Eq.~\eqref{eq:unitaryL} by imposition
	of the Galileon symmetry, after taking advantage of possible
	field redefinitions and integration by parts.
	Therefore,
	one might have some reservations that
	Eq.~\eqref{eq:pi-action} is in conflict with the conclusions of
	Cheung {\etal}, who found that the coefficient of the operator
	$\dot{\pi} ( \partial \pi )^2$ was fixed by
	$M_2(t)$.
	In a generic model this coefficient is also responsible for
	a nontrivial speed of sound, $c_s < 1$.
	Therefore the coefficient of
	$\dot{\pi} ( \partial \pi )^2$ is fixed once the background
	evolution and $c_s$ have been specified.
	On the other hand, Eqs.~\eqref{eq:pi-action}--\eqref{eq:gone-def} show that
	the coefficient of $\dot{\pi} (\partial \pi)^2$ in the Galileon
	theory is independent.
	
	This apparent discrepancy disappears if one
	accounts for all terms in Eq.~\eqref{eq:unitaryL},
	which in principle contains fourteen free coefficient functions.
	Of these, the Planck mass is measured by terrestrial experiments
	and the pair $\{ c(t), \Lambda(t) \}$ must be chosen to match
	the expansion history $H(t)$, leaving eleven free coefficients overall.
	Contributions to $\dot{\pi}(\partial \pi)^2$ arise from many of these
	operators, which are made relevant owing to the symmetries of
	the Galileon theory. (See \S\ref{sec:pdp}, where the relative
	magnitude of each term is clearly expressed
	by their contributions to the observable parameter
	$\fnl$.) It is these additional terms which break the
	expected correlation between $g_3$ and $c_s^2$.

	\subsection{Primordial density perturbations}
	\label{sec:pdp}
	
	We now proceed to compute the form of the bispectrum for
	Galileon inflation. As we have explained above,
	we focus on nongaussianities because we expect them
	to parametrize the
	difference between inequivalent choices of the
	inflationary Lagrangian.
	Although it is also important to study the properties of
	two-point statistics, these are effectively constrained to match
	observation by the closeness of the background solution to
	a de Sitter era with slowly varying $H$.
	
	When computed using Eq.~\eqref{eq:pi-action},
	the correlation properties of the primordial density perturbation
	can be expressed in terms of the coefficients of the relevant
	operators, which are $\alpha$, $c_s$, $g_1$, $g_3$ and $g_4$.
	These represent the largest contribution to each observable.
	We have argued that there are two irrelevant operators,
	$\lambda^3 \phi$ and $m^2 \phi^2$,
	which although not invariant under the
	Galilean symmetry can self-consistently be made small,
	but play an essential role in ending inflation.
	These correct the predictions of
	Eq.~\eqref{eq:pi-action}. Their influence can be
	accommodated by
	inclusion of the first subleading slow-roll corrections.
	As remarked by
	Kobayashi {\etal} \cite{Kobayashi:2010cm}, another
	reason to calculate these corrections is that $c_s$
	can exhibit a modest but non-negligible
	variation over the duration of inflation,
	which manifests itself as a correction which is formally of
	first subleading order.
	
	Corrections of this kind were calculated for the two-point function
	in canonical inflation by Stewart \& Lyth
	\cite{Stewart:1993bc}.
	The two-point function describing correlations induced
	by the quadratic part of Eq.~\eqref{eq:pi-action}
	was given at leading order by Garriga \& Mukhanov \cite{Garriga:1999vw}.
	Subleading slow-roll corrections
	were later obtained
	for the case $\LagM = P(X,\phi)$
	by Chen {\etal} \cite{Chen:2006nt},
	who also gave expressions for slow-roll corrections
	to the bispectrum. These involved quadratures of the
	sine and cosine integrals $\Si x$ and $\Ci x$ which could not
	be evaluated in closed form.
	Very recently, Kobayashi {\etal} \cite{Kobayashi:2010cm}
	obtained slow-roll corrections to the
	two-point function in models of the form
	$\LagM = P(X,\phi) + F(X,\phi) \Box \phi$, which includes the
	term involving $c_3$ in Eq.~\eqref{eq:galileon} as a special case
	but not the terms containing $c_4$ and $c_5$.
	
	\para{Two-point correlations}
	We define $s = H \dot{c}_s / c_s$
	\cite{Seery:2005wm},
	which measures the time dependence of the speed of sound.
	Similarly we shall require the rate of variation per e-fold
	of each time-dependent coefficient
	in Eq.~\eqref{eq:pi-action}.
	We define
	\begin{align}
		v & = \frac{\dot{\alpha}}{H \alpha} \\
		h_i & = \frac{\dot{g}_i}{H g_i}
	\end{align}
	for $i = 1$, $3$ and $4$,
	and treat all these combinations as the same order of magnitude
	as $\epsilon$ and $\eta$.
	Certain combinations of these parameters occur frequently, for which
	it is convenient to define abbreviations,
	\begin{align}
		\label{eq:lambda}
		\lambda & = \epsilon + \frac{v}{2} + \frac{3 s}{2} \\
		\label{eq:muzero}
		\mu_0 & = \epsilon + v + 2s + \im \frac{\pi}{2} \lambda \\
		\label{eq:muone}
		\mu_1 & = \epsilon + s - \im \frac{\pi}{2} \lambda .
	\end{align}
	We note that the combination $\mu_0 + \mu_1 = 2 \lambda$ is purely real.
	
	Translating to conformal time, defined by
	$\tau = \int_\infty^t \d t / a(t)$,
	the two-point function can be written
	\begin{equation}
		\langle \pi( \vect{k}_1, \tau ) \pi (\vect{k}_2, \tau' ) \rangle
		= (2\pi)^3 \delta(\vect{k}_1 + \vect{k}_2) G_{k_1}(\tau, \tau')
		\label{eq:twopt-def}
	\end{equation}
	in which $G_k$ is defined by
	\begin{equation}
		G_k = \left\{
			\begin{array}{l@{\hspace{3mm}}l}
				u_k^\ast(\tau') u_k(\tau) & \tau < \tau' \\
				u_k^\ast(\tau) u_k(\tau') & \tau' < \tau
			\end{array}
		\right.
		\; .
		\label{eq:green-def}
	\end{equation}
	The elementary wavefunction $u_k$ out of which this two-point function
	is built satisfies
	\begin{equation}
		u_k(\tau) =
		\frac{\sqrt{\pi}}{2\sqrt{2}}
		\frac{(1+s)^{1/2}}{a(\tau)}
		\sqrt{\frac{\tau}{\alpha(\tau)}}
		H_{\nu}^{(2)}[ - k c_s (1+s) \tau ] ,
		\label{eq:u-def}
	\end{equation}
	where $H_{\nu}^{(2)}$ is the Hankel function of the second kind.
	The order can be written $\nu = 3/2 + \lambda$
		\footnote{These slow-roll combinations differ slightly from
		Eqs.~(A.9)--(A.10) of Ref.~\cite{Chen:2006nt},
		because we are expressing each Hankel function in terms
		of the combination $- k c_s \tau$, whereas the authors
		of Ref.~\cite{Chen:2006nt} used the combination
		$k c_s / aH$. These differ at first order in slow-roll
		parameters, but our expressions are equivalent.}.
	
	The power spectrum, $P(k, \tau)$,
	is defined by the equal-time correlation function,
	\begin{equation}
		P(k, \tau) = G_{k}(\tau, \tau) .
		\label{eq:ps-def}
	\end{equation}
	On superhorizon scales, where
	$|k \tau| \ll 1$,
	Eqs.~\eqref{eq:twopt-def}--\eqref{eq:ps-def}
	imply that the power spectrum achieves a time-independent value
	\begin{equation}
		P(k, \tau) \rightarrow P(k) =
		\frac{H_{k}^2}{2(k c_{s\mid k})^3}
		\frac{1 + 2 E_k}{2 \alpha_k} ,
		\label{eq:full-ps}
	\end{equation}
	where $E$ is the combination
	\begin{equation}
		E = - \epsilon - s + ( 2 - \EulerGamma - \ln 2 ) \lambda ,
		\label{eq:E-def}
	\end{equation}
	in which $\EulerGamma$ is the Euler--Mascheroni constant,
	and a subscript $k$ denotes evaluation at the time
	$k c_s \tau = -1$.
	Eq.~\eqref{eq:full-ps}
	has engineering dimension $[\text{mass}]^{-5}$, showing
	that the two-point function,
	Eq.~\eqref{eq:twopt-def}, has engineering dimension $[\text{mass}]^{-8}$
	which is correct given our definition of $\pi$.
	The time-independence of $P(k)$ outside the horizon can be understood
	as a consequence of the shift symmetry of
	Eq.~\eqref{eq:action}, which guaranteed the conservation of
	$\zeta$ outside the horizon
		\footnote{If $\zeta = H \pi$ then only one of $\zeta$ or $\pi$
		can be constant if $H$ is changing.
		Since it is $\zeta$ which must be constant,
		it may seem puzzling
		that we find the power spectrum of $\pi$ to be time
		independent.
		The resolution is that the time-dependence of $H$
		is generated by the irrelevant operators
		$\phi$ and $\phi^2$ which were introduced to softly
		break the shift symmetry of Eq.~\eqref{eq:action}.
		Together with the couplings to gravity which have been
		discarded, these generate a mass term
		which sources time evolution of $\pi$.
		Their influence has not been included in
		our calculation of the correlation function,
		which does not capture the slow time dependence they generate in
		$\pi$.
		To obtain an accurate estimate of the true conserved
		value of $\zeta$ we should set $\zeta = H_k \pi$
		at horizon crossing.}.
	
	In writing expressions valid to subleading order in slow-roll
	parameters we must be cautious when specifying the time at which
	background quantities such as $H$, $c_s$ and the slow-roll
	parameters are to be evaluated. A small shift
	in the evaluation point translates to a difference of
	$\Or(1)$ in the slow-roll suppressed terms.
	We therefore have some freedom to rearrange coefficients
	by judicious choice of the time of evaluation.

	\para{Wavefunction corrections}
	At the level of the three-point function, slow-roll corrections arise
	from several sources. These were identified in Ref.~\cite{Chen:2006nt}.
	There are corrections due to the presence of time-dependent,
	slow-roll suppressed factors at each vertex.
	In addition there are corrections involving the combination $E$
	which arise from taking the $|k\tau| \ll 1$ limit of each external line.
	A third class of corrections arise from modifications to the
	wavefunctions carried by internal lines.
	These were obtained explicitly by Chen {\etal} \cite{Chen:2006nt},
	but in this section we give a slightly different treatment which will
	enable us to evaluate the final three-point function in closed form.
	Further details are given in Appendix \ref{appendix:ei-integrals}.
	
	\eject

	The background wavefunctions can be written
	\begin{equation}
		u(k, \tau) =
		\frac{\im H_k}{2 \sqrt{\alpha_k}}
		\frac{1}{(k c_{s\mid k})^{3/2}}
		(1 - \im k c_{s\mid k} \tau) \e{\im k c_{s\mid k} \tau} .
	\end{equation}
	The $\Or(\epsilon)$ correction, $\delta u(k, \tau)$, is obtained by
	expanding Eq.~\eqref{eq:u-def} uniformly to first order in small
	quantities.
	The variation
	with respect to the order, $\nu$, of each Hankel function can be
	evaluated using expressions (B.42)--(B.46) of Ref.~\cite{Chen:2006nt}.
	Assembling sine and cosine integral terms, these expressions can be
	rewritten to find
	\begin{equation}
	\begin{split}
		& \frac{\partial H_{\nu}^{(2)}(x)}{\partial \nu}
		= - \frac{\im}{x^{3/2}}
		\sqrt{\frac{2}{\pi}}
		\\ & \hspace{-1.5mm}\mbox{} \times
		\left[
			\e{\im x}(1-\im x) \Ei(-2\im x)
			- 2 \e{-\im x}
			- \im \frac{\pi}{2} \e{-\im x}(1+\im x)
		\right] .
	\end{split}
	\label{eq:nu-vary}
	\end{equation}

	Using Eq.~\eqref{eq:nu-vary} and Eq.~\eqref{eq:u-def}, and rotating
	the contour of integration of $\Ei$, we finally obtain
	\begin{widetext}
	\begin{equation}
	\begin{split}
		\delta u(k, \tau) =
		\frac{\im H_k}{2\sqrt{\alpha_k}}
		\frac{1}{(k c_{s\mid k})^{3/2}}
		& \Bigg\{
			- \lambda_k \e{-\im k c_{s\mid k} \tau} (1+\im k c_{s\mid k} \tau)
			\int_{-\infty}^{\tau} \frac{\d \xi}{\xi} \;
			\e{2 \im k c_{s\mid k} \xi}
			\\ & +
			\e{\im k c_{s\mid k} \tau}
			\left[
				\mu_{0\mid k} + \im \mu_{1\mid k} k c_{s\mid k} \tau + s_k
				k^2 c_{s\mid k}^2 \tau^2
				+ \lambda_k N_k - \im \lambda_k k c_{s\mid k} N_k \tau
				- s_k k^2 c_{s\mid k}^2 N \tau^2
			\right]
		\Bigg\}
	\end{split}
	\label{eq:delta-u}
	\end{equation}
	\end{widetext}
	We have defined $N_k = \ln | k c_{s\mid k} \tau |$.
	The remaining integral is to be taken over a contour
	displaced slightly above the negative real axis for large $|\xi|$,
	which renders it finite.
	Similar integrals are generated in the Schwinger (or ``in--in'')
	formulation of quantum field theory, where the same contour prescription
	is obtained after accounting for $\im \epsilon$ terms
	which project onto the vacuum at past infinity
	\cite{Maldacena:2002vr,Weinberg:2005vy}.
	Differentiating with respect to $\tau$
	and using Eqs.~\eqref{eq:lambda}--\eqref{eq:muone} one can confirm
	that,
	despite the appearance of an apparent logarithmic singularity,
	$\delta u(k,\tau)' \rightarrow 0$ in the limit $\tau \rightarrow 0$.
	This is the same behaviour as $u(k,\tau)$ itself, and guarantees that
	the introduction of slow-roll corrections does not cause a
	convergent time integral to become divergent.

	Log-divergent integrals of the form
	appearing in Eq.~\eqref{eq:delta-u} have previously been obtained in
	Refs.~\cite{Seery:2008ax,Seery:2008ms,Adshead:2009cb},
	which discussed the possibility of singularities for
	certain kinematic configurations of the $k_i$,
	including the `squeezed' configurations where one $k_i$ becomes
	much smaller than the other two.
	The behaviour of the three-point function in this limit is not
	trivial, but it can be determined nonperturbatively
	in the single-field framework following an argument due to Maldacena
	\cite{Maldacena:2002vr,Creminelli:2004yq,*Cheung:2007sv},
	and is known to be regular.
	Therefore any singularities arising from this log-divergent integral
	must cancel.
	We have confirmed
	that our final expressions contain no singularities, but we discuss
	the significance of these potential divergences
	in Appendix~\ref{appendix:ei-integrals}.
	
	\para{Three-point correlations}
	We give technical details of the calculation of the three-point functions
	arising from each cubic operator in Eq.~\eqref{eq:pi-action}
	in Appendix~\ref{appendix:threept}.
	In this section we report the final values of
	$\fnl$, specialized to the equilateral limit
	where all $k_i$ have a common magnitude.
	We would typically expect the bispectrum to be maximized on
	a configuration close to equilateral, and this
	limit should give a good estimate of the magnitude of the
	bispectrum on this peak configuration.
	
	We adopt the convention that background quantities
	are to be evaluated at the horizon-crossing time
	corresponding to the symmetric point
	$k_t = k_1 + k_2 + k_3$,
	and denote evaluation at this time by a subscript `$\star$'.
	This is somewhat larger than any individual $k_i$.
	In the equilateral case this moves the point of evaluation to
	$\ln 3 \approx 1$ e-folds after the common time of horizon exit.
	
	We define the bispectrum $B_\tau(k_1, k_2, k_3)$ by
	\begin{equation}
	\begin{split}
		\langle \pi (\vect{k}_1, \tau) & \pi(\vect{k}_2, \tau)
		\pi(\vect{k}_3, \tau) \rangle
		=
		\\ &
		(2\pi)^3 \delta(\vect{k}_1 + \vect{k}_2 + \vect{k}_3)
		B_\tau(k_1, k_2, k_3)
	\end{split}
	\label{eq:bispectrum-def}
	\end{equation}
	where the momentum-conservation condition allows us to make
	$B$ a function of the magnitudes $k_i$ alone, independent of the
	relative orientation among the $\vect{k}_i$.
	We define $\fnl$ to be the reduced bispectrum
	\begin{equation}
	\begin{split}
		B(& k_1, k_2, k_3) =
		\frac{6}{5} \fnl
		\\ &	\times
		\Big[
			P(k_1) P(k_2) + P(k_1) P(k_3) + P(k_2) P(k_3)
		\Big] ,
	\end{split}
	\label{eq:fnl-def}
	\end{equation}
	where all quantities are evaluated at time $\tau$.
	Our convention that the background quantities in
	each copy of the power spectrum $P(k)$ are to be evaluated at
	$\tau_\star$ implies that there is a logarithmic correction
	to Eq.~\eqref{eq:full-ps} proportional to $\ln k/k_t$.
	Combining Eqs.~\eqref{eq:full-ps} and~\eqref{eq:fnl-def}
	we obtain
	\begin{equation}
	\begin{split}
		\frac{6}{5} \fnl =
		&
		\frac{\prod_i k_i^3}{\sum_i k_i^3 ( 1 + 4 E_\star - 2
		\lambda_\star \ln [ k_i^{-1} k_t^{-2} \prod_j k_j ] )}
		\\ & \mbox{} \times
		\left( \frac{H_\star^2}{4\alpha_\star c_{s\star}^3} \right)^{-2}
		B(k_1, k_2, k_3) .
	\end{split}
	\end{equation}
	To obtain our final answers we expand this
	expression uniformly to first order in quantities of
	$\Or(\epsilon)$.
	We define a numerical constant $\Cconst$,
	satisfying $\coth \Cconst = 5$.
	In the equilateral limit $k_i = k$ for all $i$,
	each $\fnl$ becomes independent of $k$ for dimensional reasons and
	we find
	\begin{widetext}
	\begin{align}
		\nonumber
		\frac{6}{5} \fnl^{\dot{\pi}^3}
		& = \frac{2}{27} \frac{g_{1\star} H_\star}{\alpha_\star}
			\Bigg( 1 + \frac{2 \EulerGamma - 3}{2} h_{1\star}
				+ \frac{160 \Cconst - 2 \EulerGamma - 29}{2} v_\star
				+ \frac{480 \Cconst - 98}{2} s_\star
				+ \frac{320 \Cconst - 2 \EulerGamma - 63}{2} \epsilon_\star
			\Bigg) \\
		& = \frac{2}{27} \frac{g_{1\star} H_\star}{\alpha_\star}
			\left( 1 - 0.923 h_{1\star} + 1.141 v_\star
			- 0.344 s_\star + 0.360 \epsilon_\star \right)
		\label{eq:fnl-a}
		\\ \nonumber
		\frac{6}{5} \fnl^{\dot{\pi} (\partial \pi)^2}
		& = - \frac{17}{54 c^2_{s\star}} \frac{g_{3\star} H_\star}{\alpha_\star}
			\Bigg( 1 + \frac{17 \EulerGamma - 9}{17} h_{3\star}
				+ \frac{64 \omega - 17 \EulerGamma + 22}{17} v_\star
				+ \frac{192 \omega - 34 \EulerGamma + 40}{17} s_\star
				\\ \nonumber & \hspace{3.1cm} \mbox{}
				+ \frac{128 \omega - 17 \EulerGamma + 18}{17}
					\epsilon_\star
			\Bigg) \\
		& = - \frac{17}{54 c_{s\star}^2} \frac{g_{3\star} H_\star}{\alpha_\star}
			\left( 1 + 0.048 h_{3\star} + 1.480 v_\star
				+ 3.489 s_\star + 2.008 \epsilon_\star \right)
		\label{eq:fnl-b}
		\\ \nonumber
		\frac{6}{5} \fnl^{\partial^2 \pi (\partial \pi)^2}
		& = - \frac{13}{27 c_{s\star}^4} \frac{g_{4\star} H_\star^2}
			{\alpha_\star}
			\Bigg( 1 + \frac{6 \EulerGamma - 5}{6} h_{4\star}
				+ \frac{173 - 26 \EulerGamma - 512 \omega}{26} v_\star
				+ \frac{733 - 156 \EulerGamma - 2304 \omega}{39}
					s_\star
				\\ \nonumber & \hspace{3.1cm} \mbox{}
				+ \frac{147 - 26 \EulerGamma - 512 \omega}{13}
					\epsilon_\star
			\Bigg) \\
		& = - \frac{13}{27 c_{s\star}^4} \frac{g_{4\star} H_\star^2}
			{\alpha_\star}
			\left( 1 - 0.256 h_{4\star} + 2.084 v_\star
				+ 4.509 s_\star + 2.169 \epsilon_\star \right) .
		\label{eq:fnl-c}
	\end{align}
	\end{widetext}
	On the other hand,
	in the `squeezed' limit where one $k_i$ becomes much less than the other
	two, each $\fnl$ decays to zero.
	Expressions for $\fnl$
	which describe the complete momentum dependence
	can be extracted from the three-point functions given in
	Appendix~\ref{appendix:threept}, but because they are lengthy
	and ultimately not illuminating we do not write them explicitly.
	
	The leading-order contributions from each
	of these operators were recently computed
	by Mizuno \& Koyama \cite{Mizuno:2010ag}.
	We have verified that the leading terms of
	Eqs.~\eqref{eq:c3ba}, \eqref{eq:c4ba} and~\eqref{eq:c-a},
	which generate the leading-order
	terms of Eqs.~\eqref{eq:fnl-a}--\eqref{eq:fnl-c},
	correspond with Eqs.~(32)--(36) of
	Mizuno \& Koyama.
	
	Note that, although we are using the conventional notation
	`$\fnl$' to denote the reduced bispectrum,
	Eqs.~\eqref{eq:fnl-a}--\eqref{eq:fnl-c}
	are not directly measurable quantities, because
	\eqref{eq:bispectrum-def}--\eqref{eq:fnl-def}
	define them in terms of $\pi$.
	Indeed, according to these definitions,
	Eqs.~\eqref{eq:fnl-a}--\eqref{eq:fnl-c}
	express
	$\langle \pi^3 \rangle / \langle \pi^2 \rangle^2$,
	after removal of the momentum-conservation $\delta$-function
	from each correlator,
	for different choices of $\langle \pi^3 \rangle$.
	Therefore they possess engineering dimension
	$[\text{mass}]$.
	The observable quantity is
	the ratio $\langle \zeta^3 \rangle / \langle \zeta^2 \rangle^2$.
	The appropriate $\fnl$ which measure this ratio are obtained from
	Eqs.~\eqref{eq:fnl-a}--\eqref{eq:fnl-c} after division by $H_\star$,
	and are dimensionless.
	
	\para{Dependence on $c_s^2$}
	Eqs.~\eqref{eq:fnl-a}--\eqref{eq:fnl-c} express predictions
	for the three-point correlations generated
	by the Galileon Lagrangian, assuming that the coefficient of each relevant
	operator can be determined by measurement.

	This pattern of three-point correlations gives rise to an interesting
	phenomenology, considerably broader
	than has previously been encountered using noncanonical models.
	In theories such as DBI and $k$-inflation, it is a familiar result
	that $\fnl \sim c_{s\star}^{-2}$ \cite{Chen:2006nt}.
	Eq.~\eqref{eq:fnl-c} already shows that this is not guaranteed
	in a model exhibiting Galilean invariance,
	unless the background conspires to require $g_4/\alpha$
	proportional to $c_s^2$,
	and as we will argue below
	this is not automatically the case.
	This effect is visible in the calculation of
	Mizuno \& Koyama, although these authors did not discuss its
	significance.
	
	The possibility that $\fnl$
	is \emph{not} proportional to $c_{s\star}^{-2}$
	in the limit of small sound speed
	has not previously
	been noticed. Why is this?
	Cheung {\etal} observed that the leading relevant operator
	contributing to the three-point function would be
	$\dot{\pi}(\partial \pi)^2$, since this is suppressed by fewest
	gradients.
	In a generic theory where each operator enters with an
	approximate common mass scale $M$,
	the operator $\dot{\pi} (\partial \pi)^2$ is dimension six after
	canonical normalization, whereas $\partial^2 \pi (\partial \pi)^2$
	is dimension seven. Accordingly we would expect
	$\dot{\pi} (\partial \pi)^2$
	to dominate $\fnl$ at wavenumbers
	$k \lesssim M$.
	The crucial point is that,
	if $\dot{\pi} (\partial \pi)^2$ is the only relevant cubic operator,
	then it arises from a term which also fixes the speed of sound
	\cite{Cheung:2007st}.
	As a result, the nonlinearly realized Lorentz invariance requires
	$\fnl \sim c_{s\star}^{-2}$.
	
	The important point we wish to emphasize
	is that this is
	not forced to occur in Eq.~\eqref{eq:fnl-c}, because
	it is a dimension-seven operator rather than
	$\dot{\pi}(\partial \pi)^2$ which gives this contribution.
	We will discuss this in more detail below.
	One might wonder whether even more powers of $c_{s \star}^2$ can
	accumulate in the denominator.
	However, this is not possible.
	The factor $c_{s\star}^{-4}$ arises from an operator which is
	not suppressed by any temporal gradients, each of which
	contributes a factor $c_{s\star}^2$.
	No matter how many spatial gradients are added, no
	greater enhancement is possible at small $c_{s\star}$.
	Before drawing any conclusions about the scaling of $\fnl$
	with $c_{s\star}$, however, it is necessary to specify
	the unknown coefficients $g_{i \star}$ and $\alpha_\star$,
	and for this one must work with a concrete model.

	Mizuno \& Koyama discussed two such models,
	each of which could be written in the generic
	form $\LagM = P(X,\phi) + F(X,\phi) \Box \phi$,
	including an approximation to the DBI Galileon example introduced in
	Ref.~\cite{deRham:2010eu}.
	They studied this example in two limits,
	differentiated by $b_D \ll 1$ and $b_D \gg 1$ in their notation,
	respectively corresponding to the dimension-six and dimension-seven
	operators dominating the cubic interactions.
	In each of these extreme limits they found
	$\fnl \sim c_{s\star}^{-2}$,
	because in these limits the coefficient of the
	dominant operator also determines $c_s^2$.
	In the limit where the dimension-six operator
	is dominant, the dimension-seven operator
	gives a contribution to $\fnl$ of order
	$c_{s\star}^{-5}$ which is reproduced by our formulas.
	However, on its own this is of limited interest because
	this contribution must be subdominant overall.
	
	The full, covariant Galileon result extends this
	in an interesting way.
	First, it shows clearly that
	it is not necessary to assume that one operator or the other
	is dominant.
	As we will explain,
	the Galileon symmetry makes it natural for them to be of
	equal importance.
	
	Second, the most general Galileon model has three
	higher-derivative interactions,
	each of which contributes to the speed of sound, $c_s$.
	If any one of the $c_3$, $c_4$ or $c_5$ operators dominates
	$\alpha$ and $\beta$, then $c_s^2$ reduces to a simple rational fraction
	which is not especially small.
	If the same operator dominates the $g_i$, then each $\fnl$
	in Eqs.~\eqref{eq:fnl-a}--\eqref{eq:fnl-c}
	reduces to $\sim H$,
	giving an $\fnl$ for $\zeta$ of order unity.
	To obtain a very small speed of sound, one must suppose that
	the $c_i$ are arranged in such a way that $\beta$ becomes small
	relative to $\alpha$.
	Having done so,
	in the absence of other constraints, there is
	still sufficient freedom
	to balance the $g_i$ in such a way that the dominant contribution
	to $\fnl$ scales parametrically faster than $c_{s\star}^{-2}$
	in the limit of small sound speed.
	This is one of the key results of this paper.

	\para{Relevance of dimension-seven operator}
	Let us explain in more detail how the dimension-six and
	dimension-seven operators are naturally of comparable importance.
	
	First consider a generic theory, without a Galilean symmetry.
	If we tune $k \sim M$, then
	a window exists in which
	the contribution from both operators may be competitive,
	but in this window we are close to the cutoff of the
	effective theory unless it is forced to higher scales by a
	Vainshtein-like effect \cite{Vainshtein:1972sx}.
	In practice, if $c_s$ is very small then the
	contribution from $\partial^2 \pi (\partial \pi)^2$ could receive
	an additional enhancement, which might make it competitive
	with $\dot{\pi} (\partial \pi)^2$. However,
	$c_s$ cannot be made too small without encountering undesirable
	stability problems
	\cite{Cheung:2007st,Bartolo:2010bu}.
	Therefore the dimension-seven operator seems effectively
	irrelevant.

	The situation changes if the Lagrangian exhibits a Galilean
	invariance.
	In this case,
	the dimension-seven operator gives a significant contribution
	because the dimension-six operators explicitly break
	the Galilean symmetry. Thus, they arise only because there is a
	nontrivial cosmological background which breaks the symmetry,
	as discussed below Eq.~\eqref{eq:galileon}.
	Because this breaking is done by the background,
	it is suppressed by powers of $H/\Lambda$. This makes the
	dimension-six operators formally of the same order as the
	dimension-seven ones.
	This possibility has not been covered in
	previous discussions of the effective field theory of inflationary
	perturbations.

	To see this in detail,
	Eqs.~\eqref{eq:fone}--\eqref{eq:ffour}
	show that $\dot{\pi} (\grad \pi)^2$ and $\partial^2 \pi
	(\partial \pi)^2$ do not enter with a common mass scale.
	Instead, the coefficient of $\dot{\pi}(\partial \pi)^2$
	is suppressed by $H$, which is the same order of magnitude
	as the extra gradient carried by
	$\partial^2 \pi (\partial \pi)^2$.
	Therefore, the
	covariant Galileon model
	(specialized to de Sitter space)
	tunes these operators to give
	precisely competitive contributions
	when $c_s \sim 1$.
	In terms of observables,
	the same conclusion is easy to obtain
	from Eqs.~\eqref{eq:fnl-a}--\eqref{eq:fnl-c},
	in which the dimension-seven contribution from
	Eq.~\eqref{eq:fnl-c} is
	na\"{\i}vely suppressed by $\Or(H)$
	in comparison with the dimension-six contributions from
	Eqs.~\eqref{eq:fnl-a}--\eqref{eq:fnl-b}.
	However, the Galileon symmetry forces
	$g_{1\star}$ and $g_{3\star}$ to contain an extra
	power of $H_\star$ [\emph{cf}. Eqs.~\eqref{eq:fone} and~\eqref{eq:fthree}],
	making the contribution to $\fnl$
	from each of Eqs.~\eqref{eq:fnl-a}--\eqref{eq:fnl-c}
	precisely comparable.
	
	From the point of view of an arbitrary effective field theory,
	such tuning would be highly unnatural.
	It is remarkable that
	in a Galileon theory this tuning is automatic,
	stable and technically natural:
	this also follows from the dimension-six operators' violation
	of the Galilean
	symmetry. Therefore we may choose them to be small
	while preserving technical naturalness, in the precise sense
	of 't Hooft \cite{'tHooft:1979bh}.
	
	\section{Conclusions}

	Successful inflationary models must obey an approximate shift symmetry
	in order to generate sufficient e-folds of inflation. This shift
	symmetry allows us to add any scalar constructed from gradients of the
	inflaton field to the inflationary Lagrangian. However, when adding
	these gradient terms we must be wary of a number of pitfalls. Adding
	arbitrary higher derivative operators to the inflationary Lagrangian
	can lead to a loss of unitarity, due to the appearance of ghost
	states. The functional form of the Lagrangian need not
	be protected from
	large renormalizations, and if the Lagrangian requires more input
	parameters than can be measured
	then the theory loses all predictivity.

	In this article we have presented a model,
	termed Galileon inflation,
	which avoids these pitfalls. Building on the success of the
	Galileon models
	of dark energy, we
	have constructed an inflationary Lagrangian containing
	noncanonical derivative operators whose form is protected
	by the covariant generalisation of the Galileon shift symmetry.	 It
	contains a finite number of operators and gives rise to second order
	field equations, implying the absence of ghosts.

	We have constructed
	the action which describes fluctuations about the Galileon inflationary
	solution to third order in perturbations.	In
	contrast to previous claims in the literature we find that none of the
	coefficients of the terms at third order in the inflationary
	fluctuation are fixed by matching to
	the background evolution and the two-point statistics.
	Therefore Galileon
	inflation is an explicit example of a new class of higher
	derivative inflationary models where the nongaussianity is not
	constrained to obey $\fnl\sim 1/c_s^2$.	

	\begin{acknowledgments}
		DS would like to thank
		the Theory Group at the Deutsches Elektronen-Synchrotron DESY
		for their hospitality.
		CB is supported by the German Science Foundation (DFG) under
		the Collaborative Research Centre (SFB) 676. CdR is funded by the SNF.
		DS was supported by the Science and Technology Facilities Council
		[grant number ST/F002858/1], and acknowledges hospitality and
		support from the Perimeter Institute of Theoretical Physics,
		where this work was initiated.
		
		We would like to thank Raquel Ribeiro for pointing out
		typos in a number of equations.
	\end{acknowledgments}
	
	\appendix
	
	\section{Integrals of the $\Ei$ function}
	\label{appendix:ei-integrals}
	
	To obtain expressions for $\fnl$ in closed form, we are obliged to
	compute several integrals over the exponential integral function
	$\Ei$,
	\begin{equation}
		\Ei x = \int_{-\infty}^x \frac{\e{t}}{t} \, \d t ,
		\label{eq:ei-def}
	\end{equation}
	defined
	for real nonzero $x$. If $x < 0$ the integral is manifestly
	well-defined;
	if $x > 0$ it must be understood as a Cauchy principal value.
	If $x = 0$ there is a logarithmic singularity which cannot be removed
	by the principal value technique, so the
	origin must explicitly be excluded from the domain of $\Ei$.
	The exponential integral is related to the sine and cosine integrals
	by analogues of Euler's formula,
	\begin{align}
		\Ci x + \im \Si x & = \Ei(\im x) + \frac{\im \pi}{2} \\
		\Ci x - \im \Si x & = \Ei(-\im x) - \frac{\im \pi}{2} .
	\end{align}
	
	In constructing the $\Or(\epsilon)$ perturbed wavefunctions in
	\S\ref{sec:flucts} we must evaluate $\Ei(2\im k c_s \tau)$,
	which can be
	defined by complexifying $t$ in Eq.~\eqref{eq:ei-def}
	and interpreting the integral over a suitable contour.
	This contour
	passes from the region where $|t| \rightarrow \infty$
	with $\pi/2 < \arg t < 3\pi/2$ and terminates at $x$.
	Because $\tau < 0$, for $x = 2\im k c_s \tau$ the
	terminal point lies on the
	negative imaginary axis.
	Application of Cauchy's theorem allows the contour of integration
	to be rotated, obtaining the result quoted
	in Eq.~\eqref{eq:delta-u}.
	
	\para{Convergent series representation}
	In calculating $\fnl$ we require integrals of the form
	\begin{equation}
		I_0(k_3) = \int_{-\infty}^\tau \d \zeta \;
			\e{\im (k_1 + k_2 - k_3) c_s \zeta}
		\int_{-\infty}^\zeta \frac{\d \xi}{\xi} \;
			\e{2\im k_3 c_s \xi} ,
		\label{eq:Ithree-def}
	\end{equation}
	where although $I_0$ is a function of $k_1$, $k_2$ and $k_3$
	we have
	adopted the convention that
	only the momentum which occurs asymmetrically is indicated
	explicitly, in this case $k_3$,
	leaving the symmetric dependence on $k_1$ and $k_2$ implicit.
	It is convenient to define variables $\vartheta_3$ and $\theta_3$,
	satisfying
	\begin{equation}
		\vartheta_3 = 1 - \frac{2 k_3}{k_t}
		\quad
		\text{and}
		\quad
		\theta_3 = \frac{1 - \vartheta_3}{2\vartheta_3} ,
	\end{equation}
	in terms of which,
	after further contour rotations,
	Eq.~\eqref{eq:Ithree-def} can be rewritten
	\begin{equation}
		I_0(k_3) = \frac{\im}{\vartheta_3 k_t c_s}
			\int_{\infty}^{0} \d u \; \e{-u} \int_{\infty}^{\theta_3 u}
			\frac{\d v}{v} \; \e{-2 v} .
		\label{eq:Ithree-rotate}
	\end{equation}
	We have taken the limit $\tau \rightarrow 0$, which incurs an
	exponentially small error if $\tau$ is sufficiently late that
	the wavenumber $k_t$ is a few e-folds outside the horizon
	\cite{Maldacena:2002vr}.
	In principle there may be an obstruction to carrying out the
	contour rotation if $\vartheta_3 = 0$, which occurs in the squeezed
	limit where $k_1$ or $k_2$ approaches zero.
	In this limit, the $\zeta$-integral in Eq.~\eqref{eq:Ithree-def}
	diverges.
	As discussed in Refs.~\cite{Seery:2008ax,Adshead:2009cb}
	one should define this limit by analytic continuation,
	first carrying out the integral with $\vartheta_3 > 0$
	and only subsequently squeezing one of the momenta to zero.
	For this reason we need not worry about subtleties associated with
	rotation of either integral to the imaginary axis.
	There is also a logarithmic singularity when $\theta_3 = 0$,
	causing the interior $v$-integral to diverge.
	
	The $v$-integral
	has a convergent
	series representation for all $\theta_3 u > 0$, which is guaranteed
	to be the case in virtue of
	our assumptions and the domain of integration for $u$.
	This convergent series representation is inherited from
	$\e{-2v}$, which is an entire function.
	We find
	\begin{equation}
		\int_{\infty}^{\theta_3 u} \frac{\d v}{v} \; \e{-2v}
		= \EulerGamma + \ln 2 \theta_3 u +
			\sum_{n=1}^\infty \frac{(-2 \theta_3 u)^n}{n \cdot n!} .
		\label{eq:clare-formula}
	\end{equation}
	Because Eq.~\eqref{eq:clare-formula} is valid for all $u > 0$ we may
	substitute in Eq.~\eqref{eq:Ithree-rotate}
	and integrate term-by-term, which yields
	\begin{equation}
		I_0(k_3) = \frac{\im}{\vartheta_3 k_t c_s} \left(
			\sum_{n=1}^{\infty} (-1)^{n+1} \frac{(2 \theta_3)^n}{n}
			- \ln 2 \theta_3 \right) .
		\label{eq:term-by-term}
	\end{equation}
	
	\para{Analytic continuation}
	The sum converges for $|\theta_3| \leq 1/2$.
	Since $I_0$ is defined by an integral it is an analytic function of
	$\theta_3$ on an open neighbourhood
	of the positive real axis, although excluding the origin
	$\theta_3 = 0$ where we have noted that Eq.~\eqref{eq:Ithree-rotate}
	exhibits a logarithmic singularity.
	Therefore $I_0$
	may be determined by analytic continuation
	of Eq.~\eqref{eq:term-by-term} from any open set where the sum
	converges.
	We conclude that $I_0$ has the compact
	representation
	\begin{equation}
		I_0(k_3) = \frac{\im}{\vartheta_3 k_t c_s} \ln \frac{1 + 2 \theta_3}
			{2 \theta_3}
			= - \frac{\im}{\vartheta_3 k_t c_s} \ln (1 - \vartheta_3) .
		\label{eq:Ithree-compact}
	\end{equation}
	Eq.~\eqref{eq:Ithree-compact} correctly
	reproduces the expected pole as $\vartheta_3 \rightarrow 0$,
	associated with a failure of convergence in the $\zeta$-integral
	of Eq.~\eqref{eq:Ithree-def}.
	It also correctly reproduces the logarithmic singularity
	as $\theta_3 \rightarrow 0$, which corresponds
	to the limit $\vartheta_3 \rightarrow 1$.
	This is \emph{also} a squeezed limit, occurring when $k_3 \rightarrow 0$,
	and forces $k_1 = k_2$.
	Away from these singular points,
	Eq.~\eqref{eq:Ithree-compact} expresses $I_0(k_3)$ as a well-defined
	analytic function of $\theta_3$.
	In practice these singularities cancel among themselves
	in our final expressions
	for $\fnl$, which serves as a consistency check on the calculation.
	Note that the equilateral limit is $\vartheta_3 = 1/3$
	or $\theta_3 = 1$.	

	Repeating these steps enables us to find representations for
	integrals analogous to Eq.~\eqref{eq:Ithree-def}, with insertions
	of arbitrary polynomials in the $\zeta$-integral,
	\begin{equation}
		\hspace{-3pt}I_m(k_3) = \int_{-\infty}^\tau\hspace{-7pt} \d \zeta \;
		\zeta^m \e{\im(k_1+k_2-k_3)c_s \zeta}
		\int_{-\infty}^\zeta \hspace{-3pt} \frac{\d \xi}{\xi} \;
		\e{2\im k_3 c_s \xi}.\hspace{-2pt}
	\end{equation}
	We find
	\begin{widetext}
	\begin{align}
		\label{eq:Ione}
		I_1(k_3) & = \frac{1}{(\vartheta_3 k_t c_s)^2} \big[
			\vartheta_3 + \ln ( 1 - \vartheta_3 ) \big] \\
		\label{eq:Itwo}
		I_2(k_3) & = \frac{\im}{(\vartheta_3 k_t c_s)^3} \big[
			\vartheta_3(2 + \vartheta_3) + 2 \ln (1-\vartheta_3) \big] \\
		\label{eq:Ithree}
		I_3(k_3) & = -\frac{1}{(\vartheta_3 k_t c_s)^4} \big[
			\vartheta_3(6 + 3 \vartheta_3 + 2 \vartheta_3^2)
			+ 6 \ln (1 - \vartheta_3) \big] \\
		\label{eq:Im}
		I_m(k_3) & = \frac{\im^{m+1}}{(\vartheta_3 k_t c_s)^{m+1}}
			\left[
				2 \theta_3 (m+1)! \;
				F \left( \left. \begin{array}{c}
					1 \;\; 1 \;\; 2+m \\ 2 \;\; 2
				\end{array} \right| -2 \theta_3 \right)
				- m! \Big\{ \EulerGamma + \ln 2 \theta_3 + \psi^{(m+1)}(0)
				\Big\}
			\right] ,
	\end{align}
	\end{widetext}
	where we have retained the convention of denoting dependence on
	the single asymmetric momenta alone.
	The transcendental functions appearing here are
	the generalized
	hypergeometric function, $F$,
	and the polygamma function,
	$\psi^{(m)}$.
	For the purposes of this paper, we require only $I_m$ with $m \leq 3$.

	\begin{widetext}
	
	\section{Three-point functions}
	\label{appendix:threept}
	
	The rules used to obtain correlation functions from
	a Lagrangian such as~\eqref{eq:pi-action} are discussed at several
	places in the literature, to which we refer for calculational details
	\cite{Maldacena:2002vr,Weinberg:2005vy,Seery:2007we}.
	We calculate the contribution
	to $\langle \pi^3 \rangle$ induced in turn
	by each operator in Eq.~\eqref{eq:pi-action}. The possible operators
	are $\dot{\pi}^3$, $\dot{\pi}(\partial \pi)^2$ and $\partial^2 \pi
	(\partial \pi)^2$.
	In a model containing more than one of these, their contributions add
	linearly to the total three-point correlation function.
	
	\para{$\dot{\pi}^3$ operator}
	It is convenient to divide the calculation into
	\pt{a} a part containing
	the $\Or(1)$ contribution and $\Or(\epsilon)$ contributions
	from external lines and the vertex; and
	\pt{b} a part containing
	the $\Or(\epsilon)$ contributions from
	internal lines.
	We use
	the notation of Eqs.~\eqref{eq:lambda}--\eqref{eq:muone},
	\eqref{eq:E-def}, and~\eqref{eq:bispectrum-def}
	and label the bispectrum generated by parts \pt{a}
	and \pt{b} as $B^{(a)}$ and $B^{(b)}$, respectively.
	For part \pt{a} we find
	\begin{equation}
		B^{(a)} = \frac{g_{1\star}}{H_{\star}}
			\frac{H_\star^6}{4^3 \alpha_\star^3}
			\frac{24}{c_{s \star}^6}
			\frac{1}{k_t^3 \prod_i k_i}
			\left[
				1 + 3 E_\star - \lambda_\star \ln \frac{k_1 k_2 k_3}{k_t^3}
				+ (\epsilon_\star + h_{1\star}) \EulerGamma
				- \frac{1}{2} ( \epsilon_\star + 3 h_{1\star} )
			\right]
		\label{eq:c3ba}
	\end{equation}
	
	To evaluate \pt{b} it is first convenient to obtain an expression for the
	integral
	\begin{equation}
	\begin{split}
		J = \frac{1}{k_t c_{s\star}} \int_{-\infty}^\tau
			\d \left( \e{\im k_t c_s \xi} \right)
			\big[ & \
				\gamma_0
				+ \im \gamma_1 c_{s\star} \xi
				+ \gamma_2 c_{s\star}^2 \xi^2
				+ \im \gamma_3 c_{s\star}^3 \xi^3
				+ \gamma_4 c_{s\star}^4 \xi^4
		\\ & \mbox{}
				+ \delta_0 N_\star(\xi)
				+ \im \delta_1 c_{s\star} N_\star(\xi) \xi
				+ \delta_2 c_{s\star}^2 N_\star(\xi) \xi^2
				+ \im \delta_3 c_{s\star}^3 N_\star(\xi) \xi^3
				+ \delta_4 c_{s\star}^4 N_\star(\xi) \xi^4
			\big] ,
	\end{split}
	\end{equation}
	where $N_\star(\xi) = \ln|k_t c_{s\star} \xi|$.
	This is sufficiently general to encompass the $\tau$-integration
	required for each operator, not including integrations of
	the exponential integral function $\Ei$ which we treat separately
	and are discussed in Appendix~\ref{appendix:ei-integrals}.
	Carrying out the $\xi$ integral, we find
	\begin{equation}
		J = \frac{1}{k_t c_{s\star}} \Bigg[
			\gamma_0
			- \frac{\gamma_1 + \delta_1}{k_t}
			- \frac{2 \gamma_2 + 3 \delta_2}{k_t^2}
			+ \frac{6 \gamma_3 + 11 \delta_3}{k_t^3}
			+ \frac{24 \gamma_4 + 50 \delta_4}{k_t^4}
			- \left( \EulerGamma + \frac{\im \pi}{2} \right)
			\left( \delta_0
				- \frac{\delta_1}{k_t}
				- \frac{2\delta_2}{k_t^2}
				+ \frac{6\delta_3}{k_t^3}
				+ \frac{24\delta_4}{k_t^4}
			\right)
		\Bigg] .
		\label{eq:j-def}
	\end{equation}
	In terms of $J$ and the integral
	$I_2$ define by Eq.~\eqref{eq:Itwo}
	of Appendix~\ref{appendix:ei-integrals},
	the \pt{b} contribution can be written
	\begin{equation}
		B^{(b)} =
			\frac{g_{1\star}}{H_\star}
			\frac{H_\star^6}{4^3 \alpha_\star^3}
			\frac{6}{c_{s\star}^5}
			\frac{1}{\prod_i k_i}
			\big[ - J_3^{\dot{\pi}^3} + \im \lambda_\star
				c_{s \star}^2
				I_2(k_3) \big]
			+ \text{c.c.} + (k_3 \rightarrow k_2 \rightarrow k_1) ,
		\label{eq:a-interior}
	\end{equation}
	where $J_3^{\dot{\pi}^3}$ denotes~\eqref{eq:j-def} with the assignments
	$\gamma_0 = \gamma = \delta_0 = \delta_1 = 0$,
	and
	\begin{align}
		\gamma_2 & = s_\star - \mu_{1\star} \\
		\gamma_3 & = k_3 s_{\star} \\
		\delta_2 & = \lambda_\star - 2 s_{\star} \\
		\delta_3 & = -k_3 s_{\star} .
	\end{align}
	The notation `$+\text{c.c}$' denotes addition of the complex
	conjugate of the preceding term,
	and $k_3 \rightarrow k_2 \rightarrow k_1$ indicates that
	Eq.~\eqref{eq:a-interior} is to be symmetrized over the
	exchanges $k_3 \leftrightarrow k_2$ and $k_3 \leftrightarrow k_1$,
	yielding a final expression which is symmetric between the labels
	1, 2 and 3.
	
	\para{$\dot{\pi}(\partial \pi)^2$ operator}
	We divide the calculation into \pt{a} and \pt{b} parts, as above.
	The $\Or(1)$ contribution and $\Or(\epsilon)$ contributions from
	external lines and the vertex give
	\begin{equation}
	\begin{split}
		B^{(a)} = \frac{g_{3\star}}{H_\star}
		\frac{H_\star^6}{4^3 \alpha_\star^3}
		\frac{2}{c_{s\star}^7}
		\frac{k_3^2 (\vect{k}_1 \cdot \vect{k}_2)}{\prod_i k_i^3}
		\Bigg[ &\
			\frac{1}{k_t c_{s\star}} \left(
				1 + 3E_\star - \lambda_\star \ln \frac{k_1 k_2 k_3}{k_t^3}
			\right)
			\left( 1 + \frac{k_t(k_1 + k_2) + 2 k_1 k_2}{k_t^2} \right)
		\\ & \mbox{}
			+ J^{\dot{\pi}(\partial \pi)^3}_{3(a)}
		\Bigg]
		+ \text{c.c} + (k_3 \rightarrow k_2 \rightarrow k_1) ,
	\end{split}
	\label{eq:c4ba}
	\end{equation}
	where $J^{\dot{\pi}(\partial \pi)^3}_{3(a)}$ is defined by the assignments
	$\gamma_3 = \gamma_4 = \delta_3 = \delta_4 = 0$, and
	\begin{align}
		\gamma_0 & = \epsilon_\star \\
		\gamma_1 & = - \epsilon_\star(k_1 + k_2) \\
		\gamma_2 & = - \epsilon_\star k_1 k_2 \\
		\delta_0 & = - \epsilon_\star - h_{3\star} \\
		\delta_1 & = (\epsilon_\star + h_{3\star})(k_1 + k_2) \\
		\delta_2 & = (\epsilon_\star + h_{3\star}) k_1 k_2 .
	\end{align}
	The $\Or(\epsilon)$ corrections from internal lines contribute
	\begin{equation}
	\begin{split}
		B^{(b)} = \frac{g_{3\star}}{H_\star}
			\frac{H_\star^6}{4^3 \alpha_\star^3}
			\frac{2}{c_{s\star}^7}
			\frac{k_3^2 (\vect{k}_1 \cdot \vect{k}_2)}{\prod_i k_i^3}
			\Bigg[ {-\im} \lambda_\star \Big\{ &\
				I_0(k_3) - \im (k_1 + k_2) c_{s\star} I_1(k_3)
				- k_1 k_2 c_{s\star}^2 I_2(k_3)
		\\ & \mbox{} + I_0(k_1) - \im (k_2 - k_1) c_{s\star} I_1(k_1)
				+ k_1 k_2 c_{s\star}^2 I_2(k_1)
		\\ & \mbox{} + I_0(k_2) - \im (k_1 - k_2) c_{s\star} I_1(k_2)
				+ k_1 k_2 c_{s\star}^2 I_2(k_2) \Big\}
		+ J_{3(b)}^{\dot{\pi}(\partial \pi)^2} \Bigg]
		\\ & \mbox{} + \text{c.c.} + (k_3 \rightarrow k_2 \rightarrow k_1) ,
	\end{split}
	\end{equation}
	where $J_{3(b)}^{\dot{\pi}(\partial \pi)^2}$ is defined by the assignments
	$\gamma_4 = \delta_4 = 0$, and
	\begin{align}
		\gamma_0 & = 2 \mu_{0\star} - \mu_{1\star} + s_\star \\
		\gamma_1 & = s_\star k_3 + (k_1 + k_2)(2 \mu_{1\star} - \mu_{0\star}
			- s_\star) \\
		\gamma_2 & = k_1 k_2 ( 3 \mu_{1\star} - s_\star )
			+ s_\star (k_1^2 + k_2^2) + s_\star k_3 (k_1 + k_2) \\
		\gamma_3 & = - s_\star k_1 k_2 k_t \\
		\delta_0 & = 3\lambda_\star - 2 s_\star \\
		\delta_1 & = (2 s_\star - 3 \lambda_\star)(k_1 + k_2) - s_\star k_3 \\
		\delta_2 & = k_1 k_2 ( 2 s_\star - 3\lambda_\star)
			- s_\star (k_1^2 + k_2^2) - s_\star k_3(k_1 + k_2) \\
		\delta_3 & = s_\star k_1 k_2 k_t .
	\end{align}
	
	\para{$\partial^2 \pi (\partial \pi)^2$ operator}
	Applying the same procedure to the final operator, $\partial^2 \pi
	(\partial \pi)^2$, gives a contribution at $\Or(1)$ and including
	$\Or(\epsilon)$ terms from the external lines and vertex:
	\begin{equation}
	\begin{split}
		B^{(a)} = g_{4\star} \frac{H_\star^6}{4^3 \alpha_\star^3}
		\frac{4}{c_{s\star}^{10}}
		\frac{k_3^2 ( \vect{k}_1 \cdot \vect{k}_2 )}{k_t \prod_i k_i^3}
		\Bigg[ &\
			\left(
				1 + 3 E_\star - \lambda_\star \ln \frac{k_1 k_2 k_3}{k_t^3}
			\right)
			\left( 1 + \frac{3 k_1 k_2 k_3 + k_t K^2}{k_t^3} \right)
		\\ & \mbox{}
		- \frac{h_{4\star}}{2} \left\{
			1 + \frac{3 K^2}{k_t^2} + 11 \frac{k_1 k_2 k_3}{k_t^3}
			- \left( \EulerGamma + \frac{\im \pi}{2} \right)
			\left( 2 + \frac{2 K^2}{k_t^2} + \frac{6 k_1 k_2 k_3}{k_t^3}
			\right)
		\right\}
		\Bigg]
		\\ & \mbox{}
		+ \text{c.c} + (k_3 \rightarrow k_2 \rightarrow k_1) ,
	\end{split}
	\label{eq:c-a}
	\end{equation}
	where we have defined $K^2 = k_1 k_2 + k_1 k_3 + k_2 k_3 =
	\sum_{i < j} k_i k_j$.
	Finally we also require the \pt{b} contribution, from $\Or(\epsilon)$
	corrections to each internal line, which yields
	\begin{equation}
	\begin{split}
		B^{(b)} = g_{4\star} \frac{H_\star^6}{4^3 \alpha_\star^3}
			&
			\frac{2}{c_{s\star}^9}
			\frac{k_3^2 (\vect{k}_1 \cdot \vect{k}_2)}{\prod_i k_i^3}
		\\ & \hspace{-1cm}
		\Bigg[ {-\im} \lambda_\star \Big\{
			I_0(k_3)
			- \im (k_1 + k_2 - k_3) c_{s\star} I_1(k_3)
			+ (k_3 k_1 + k_3 k_2 - k_1 k_2) c_{s\star}^2 I_2(k_3)
			- \im k_1 k_2 k_3 c_{s\star}^3 I_3(k_3)
		\\ & \mbox{} + I_0(k_1)
			- \im (k_2 + k_3 - k_1) c_{s\star} I_1(k_1)
			+ (k_1 k_2 + k_1 k_3 - k_2 k_3) c_{s\star}^2 I_2(k_1)
			- \im k_1 k_2 k_3 c_{s\star}^3 I_3(k_1)
		\\ & \mbox{} + I_0(k_2)
			- \im (k_1 + k_3 - k_2) c_{s\star} I_1(k_2)
			+ (k_2 k_1 + k_2 k_3 - k_1 k_3) c_{s\star}^2 I_2(k_2)
			- \im k_1 k_2 k_3 c_{s\star}^3 I_3(k_2) \Big\}
		+ J_{3}^{\partial^2 \pi (\partial \pi)^2} \Bigg]
		\\ & \mbox{} + \text{c.c.} + (k_3 \rightarrow k_2 \rightarrow k_1) ,
	\end{split}
	\label{eq:c-b}
	\end{equation}
	where $J_{3}^{\partial^2 \pi (\partial \pi)^2}$ is defined by the
	assignments
	\begin{align}
		\gamma_0 & = 3 \mu_{0\star} \\
		\gamma_1 & = k_t ( \mu_{1\star} - 2 \mu_{0\star} ) \\
		\gamma_2 & = s_\star ( k_1^2 + k_2^2 + k_3^2 ) + K^2
			(2 \mu_{1\star} - \mu_{0\star}) \\
		\gamma_3 & = - s_\star \left[
				k_1^2 (k_2 + k_3) + k_2^2 (k_1 + k_3) + k_3^2 (k_1 + k_2)
			\right] - 3 \mu_{1\star} k_1 k_2 k_3 \\
		\gamma_4 & = - s_\star k_1 k_2 k_3 k_t \\
		\delta_0 & = 3 \lambda_\star \\
		\delta_1 & = - 3\lambda_\star k_t \\
		\delta_2 & = - s_\star (k_1^2 + k_2^2 + k_3^2) - 3 \lambda_\star K^2 \\
		\delta_3 & = s_\star \left[
				k_1^2 (k_2 + k_3) + k_2^2 (k_1 + k_3) + k_3^2 (k_1 + k_2)
			\right] + 3 \lambda_\star k_1 k_2 k_3 \\
		\delta_4 & = s_\star k_1 k_2 k_3 k_t .
	\end{align}
	Note that, because this interaction is symmetric apart from
	the arrangement of spatial gradients,
	Eqs.~\eqref{eq:c-a} and~\eqref{eq:c-b} are in fact
	symmetric under permutation of the labels 1, 2 and 3
	except for the overall factor $k_3^2 (\vect{k}_1 \cdot \vect{k}_2)$,
	which arises from the specific gradient combination
	appearing in $\partial^2 \pi
	(\partial \pi)^2$.
	
	\end{widetext}

\bibliography{galileon}

\end{document}